%% file: phase.tex
\documentstyle[preprint,eqsecnum,aps]{revtex}
\begin{document}
\draft
\title{Phase diagram of Regge quantum gravity coupled to SU(2) gauge
 theory}
\author{B.A.~Berg\thanks{Work supported in part by DOE under Contracts
 DE-FG05-87ER40319 and DE-FC05-85ER2500.}}
\address{Department of Physics and SCRI, Florida State University,
 Tallahassee, FL 32306, USA}
\author{W.~Beirl$^*$, B.~Krishnan\thanks{Lise-Meitner Postdoctoral Research 
 Fellow sponsored by FWF under Project M212-PHY.}, H.~Markum, and J.~Riedler}
\address{Institut f\"{u}r Kernphysik, Technische Universit\"{a}t Wien,
 A-1040 Vienna, Austria}
\date{\today}
\maketitle
\begin{abstract}
We analyze Regge quantum gravity coupled to SU(2) gauge theory on
$4^3\times 2$, $6^{3}\times 4$ and $8^{3}\times 4$ simplicial lattices.
It turns out that the window of the well-defined phase of the gravity
sector where geometrical expectation values are stable extends to 
negative gravitational couplings as well as to gauge couplings across 
the deconfinement phase transition. We study the string tension from
Polyakov loops, compare with the $\beta$-function of pure gauge theory
and conclude that a physical limit through scaling is possible.
\end{abstract}

\input{psfig}

\narrowtext
\section{Introduction}
Regge Calculus \cite{R} provides a nonperturbative way for investigations 
of Euclidean quantum gravity on a simplicial lattice and offers the 
possibility to construct a unified theory by coupling gauge fields to the 
skeleton. One remarkable finding was the discovery of an ``entropy dominated'',
well-defined phase, where the expectation values with respect to the 
pure Regge-Einstein action are stable \cite{BB1,HH1,BGM}. The next question 
addressed was the physical relevance of this regime which has been tested 
by coupling a non-Abelian gauge field to gravity \cite{BKK}. If one assumes 
that the world without gravity is described by a grand unified asymptotically 
free theory, these numerical studies investigate the relation of the 
hadronic scale to the Planck scale. In particular, it has already been
shown that confinement exists in the coupled system \cite{BBKMRpl}.
In this work, we perform a nonperturbative analysis of the phase diagram 
of Regge quantum gravity coupled to SU(2) gauge fields on several lattice
sizes in four spacetime dimensions. The stability and boundary of 
the well-defined phase is investigated on lattices of sizes up to 
$6^3\times 4$, considerably extending the previously available data.
Within this phase we find that the confinement-deconfinement
transition of conventional lattice gauge theory is still present.
We extract values for the string tension and gain some evidence
from the $\beta$-function that the window of the well-defined phase
extends to large $\beta$ values corresponding to small lattice spacings
$a$ in the order of the Planck scale.

\section{Entropy Dominated Phase}
In Regge Calculus the edge lengths are considered to be the dynamical degrees
of freedome of the discretized spacetime manifold. In $d=4$ dimensions
the geometry is Euclidean inside of a $d$-simplex and the curvature is 
concentrated at the $d-2$-subset of the lattice, the triangles. Quantization 
proceeds via the path integral although the choice of the 
gravitational measure is an unresolved issue. Investigations on regular
triangulations do not favor any of them \cite{BGM}, but for irregular
triangulations a preference for scale-invariant measures was found
\cite{BMR}. It may be that measures are divided into universality 
classes such that identical physics is obtained within one class. 

In our simulations we chose a hypercubic triangulation with 
$N_{s}^{3}\times N_{t}$ vertices and the scale-invariant measure:
\begin{equation}
D[\{l^{2}\}]\,=\,\prod_{l}\frac{dl^{2}}{l^{2}}\,, \label{eq:measure}
\end{equation}
where $l$ is used to denote the link label as well as its length. 
The system of SU(2) gauge fields coupled to quantum gravity has the 
action \cite{BKK}
\begin{equation} \label{coupact}
 S\,=\,2m_{p}^{2}\sum_{t}A_{t}\alpha_{t} - 
\frac{\beta}{2}\sum_{t}W_{t}\,{\rm Re}[ {\rm Tr}(1\,-\,U_{t})]\; , 
\label{eq:action}
\end{equation}
with the first term being the Regge-Einstein part composed of the bare 
Planck mass $m_{p}$ and $A_{t}$, $\alpha_{t}$ the area and the deficit angle 
of the triangle $t$. The addition of the SU(2) gauge term is straightforward 
and follows ordinary lattice gauge theory by assigning SU(2) matrices to the 
links. The elementary plaquettes become triangles on the Regge skeleton.
$\beta$ corresponds to the inverse gauge coupling and the weights 
\hbox{$W_{t}=\mbox{const}\times\frac{V_t}{(A_t)^2}$}, with a 4-volume $V_t$ 
assigned to each triangle, describe the coupling of gravity to the gauge 
field. $U_{t}$ is the ordered product of SU(2) matrices around $t$. In 
contrast to the flat lattice, the unit matrix in the action is important 
because the weight factors are dynamical. They are constructed such that the 
correct continuum limit is ensured in the limit of vanishing lattice spacing 
\cite{CL}.

We present Monte Carlo (MC) results concerning the boundary and stability 
of the well-defined phase. For this purpose large statistics 
simulations were performed on $4^3\times 2$ and $6^3\times 4$ lattices, 
covering a variety of $(m_p^2,\beta)$ values. The accumulated statistics
is summarized in Tables~\ref{table1} and~\ref{table2}.

\begin{figure}
\hbox{\hspace{11mm}\scriptsize{$\beta=0.0$}
      \hspace{44mm}\scriptsize{$\beta=1.5$}
      \hspace{44mm}\scriptsize{$\beta=1.6$}}
\centerline{\hbox{
 \psfig{figure=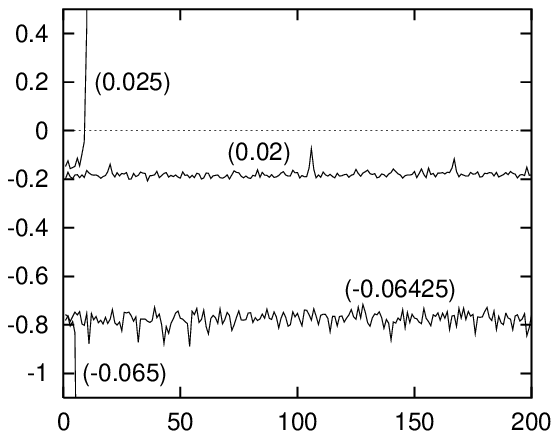,height=2in,width=2.1in}
 \psfig{figure=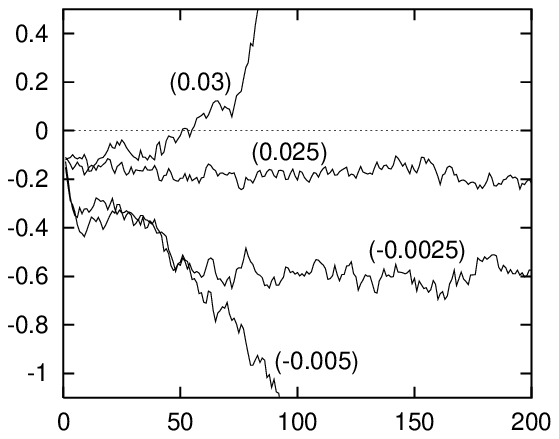,height=2in,width=2.1in}
 \psfig{figure=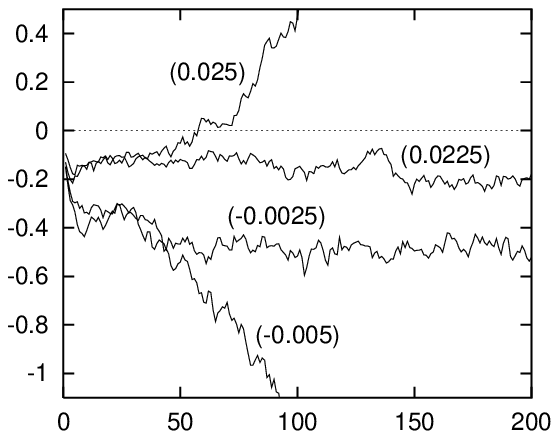,height=2in,width=2.1in}
 }}
\hbox{\hspace{11mm}\scriptsize{$\beta=0.0$}
      \hspace{44mm}\scriptsize{$\beta=1.0$}
      \hspace{44mm}\scriptsize{$\beta=1.6$}}
\centerline{\hbox{
 \psfig{figure=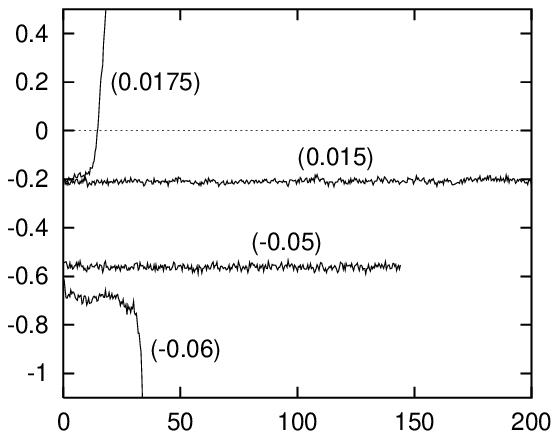,height=2in,width=2.1in}
 \psfig{figure=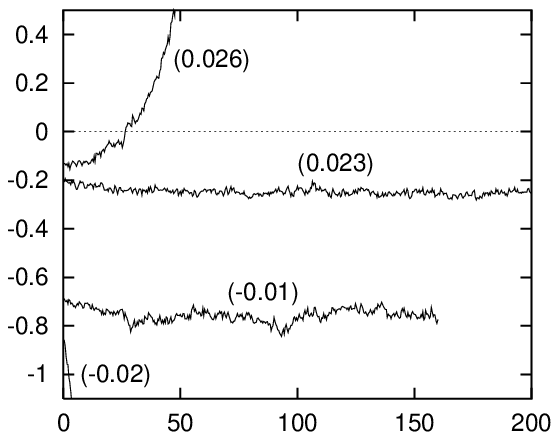,height=2in,width=2.1in}
 \psfig{figure=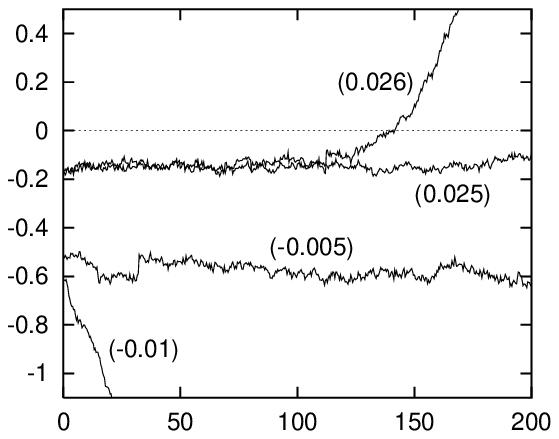,height=2in,width=2.1in}
 }}
\vspace{1cm}
\caption{\label{his} MC-time histories of the Regge action $\langle
A_t\delta_t\rangle$ for $4^3\times 2$ (upper) and $6^3\times 4$ (lower) 
lattices, with one time unit representing 1k sweeps.
The Planck masses $m_p^2$ used in the simulations are indicated in
brackets beside the curves. The plots correspond to simulations at the
$\beta$ values given in the titles.}
\end{figure}

The stability of the coupled system was analyzed from MC-time histories
of the Regge action. Examples are presented in Fig.~\ref{his}. Long
runs on the larger lattice show that it is difficult to decide whether 
the well-defined phase is stable or just metastable. In the latter case 
an additional curvature term of higher order \cite{HeHa} may stabilize
the system.

\begin{figure}
\hbox{\hspace{31mm}\scriptsize{$m_p^2=0.0225,~ \beta=1.5$}
      \hspace{25mm}\scriptsize{$m_p^2=0.0225,~ \beta=1.6$}}
\centerline{\hbox{
 \psfig{figure=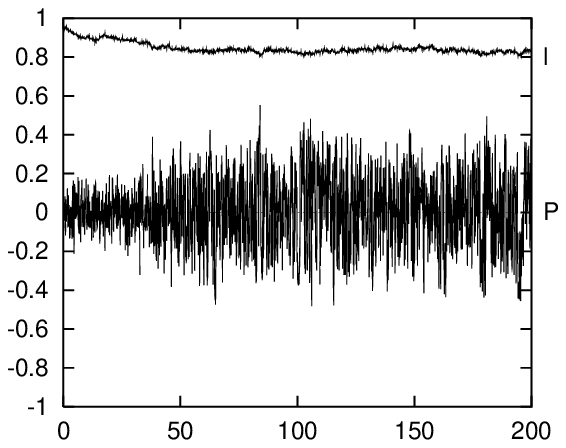,height=2in,width=2.1in}
 \psfig{figure=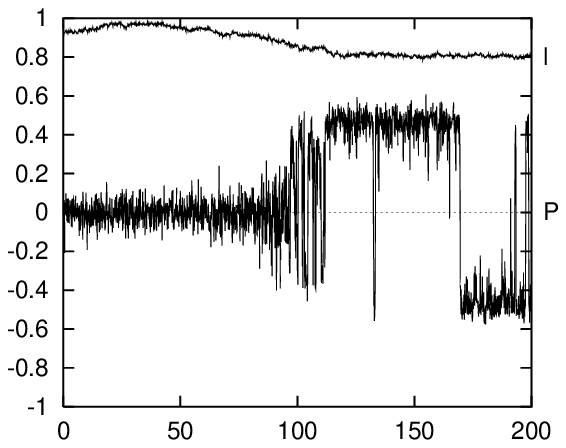,height=2in,width=2.1in}
 }}
\hbox{\hspace{31mm}\scriptsize{$m_p^2=0.020,~ \beta=1.55$}
      \hspace{25mm}\scriptsize{$m_p^2=0.023,~ \beta=1.6$}}
\centerline{\hbox{
 \psfig{figure=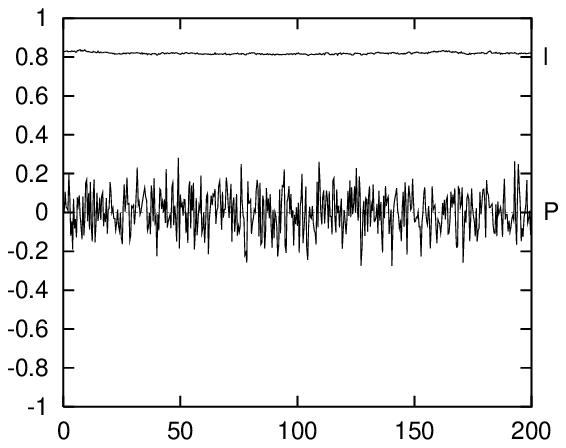,height=2in,width=2.1in}
 \psfig{figure=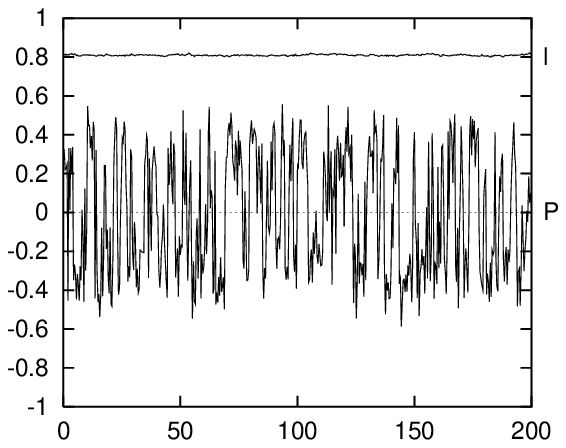,height=2in,width=2.1in}
 }}
\vspace{1cm}
\caption{\label{poli} Time histories of Polyakov loops $P$ and link lengths
$l$ on $4^3\times 2$ (upper) and $6^3\times 4$ (lower) lattices, with one 
time unit representing 1k sweeps. For reasons of comparison link lengths 
are normalized with respect to the maximum of all four runs. In the lower 
figures $P$ is multiplied by a factor of two. The $(m_p^2,\beta)$ values 
are given in the titles.}
\end{figure}

The deconfinement transition was studied from the behavior of the Polyakov
loops \hbox{$P=\frac{1}{2}\mbox{Tr}(U_1U_2\dots U_{N_t})$}. For 
$4^3\times 2$ and 
$6^3\times 4$ lattices Fig.~\ref{poli} shows MC-time histories of $P$ 
and $l$ in the confined and deconfined phase for gravitational couplings 
in the well-defined phase. The link lengths are largely independent of 
fluctuations of the order parameter. Notable is the long equilibration 
time in the deconfinement phase.

\begin{figure}
\input{phasefig3a}
\vspace{1cm}
\input{phasefig3b}
\vspace{1cm}
\caption{\label{phadia} Phase diagrams extracted from $4^3\times 2$ (a)
and $6^3\times 4$ (b) lattices. The bold vertical lines indicate the
confinement-deconfinement transition.}
\end{figure}
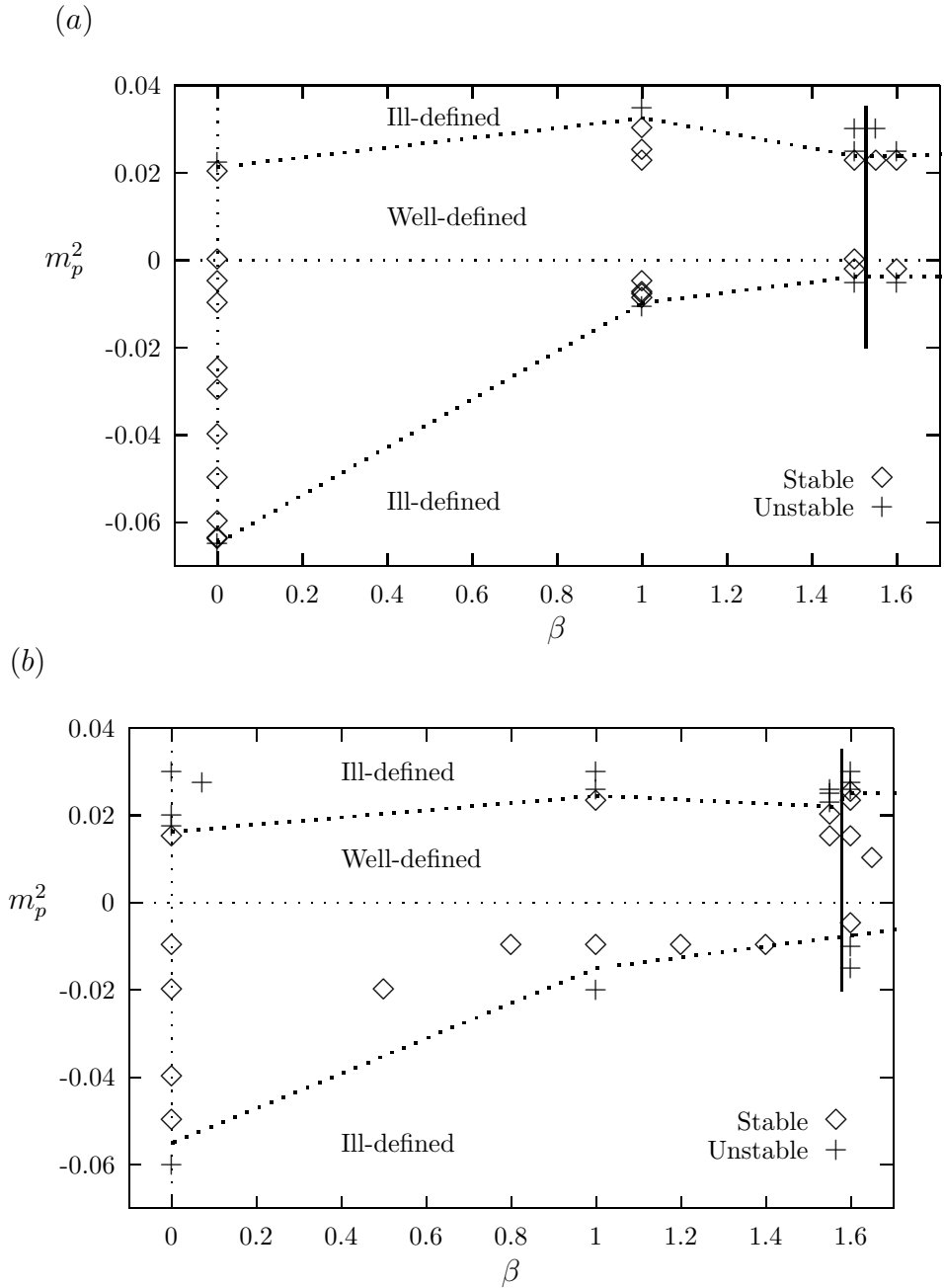

Extracted phase diagrams of the gauge-gravity system are displayed in
Fig.~\ref{phadia}. The dotted lines are to guide the  eyes and rely on the
depicted stable versus unstable data points. From the $4^3\times 2$
lattices we have numerical evidence that the stable phase extends to
$\beta=3.0$ (Table \ref{table1}). The $6^3\times 4$
lattices indicate a glitch in the well-defined--ill-defined
boundary when passing from confinement to deconfinement. However,
the present runs do not decide the question conclusively. It may be
accidental that at $\beta = 1.6$ the $m_p^2=0.025$ and $m_p^2=0.023$ 
data did not run away. For $\beta=1.55$ tunneling into the ill-defined
phase happened for these $m_p^2$ values only after more than 100k
sweeps. Our systems exhibit a small shift of the deconfinement phase
transition from $\beta_c(N_t=2)=1.525$ to $\beta_c(N_t=4)=1.575$ with
error bars less than 0.025.

\section{String Tension}
The Polyakov loop $P(R)$ in the short extent $L_{t}$ of the lattice 
describes the propagation of a static quark. We introduce a quark source and 
a sink separated by a distance $R$ and calculate the correlations of the 
Polyakov loops at these points. As in conventional SU(2) lattice gauge theory 
at finite temperature $T$ \cite{MS,KPS}, we extract from the correlation 
function
\begin{equation}
\langle P(0) P^{\dagger}(R) \rangle\,=\,\exp[- \frac{1}{T} V(R)] , 
\label{eq:correl}
\end{equation}
the quantity $V(R)$ corresponding to the potential between the static 
quark-antiquark pair. In the confinement phase $V(R)$ should 
grow linearly for large $R$ due to an infinite free energy of
isolated quarks:
\begin{equation} \label{Vc}
V_c(R)=\frac{-\alpha}{R}+\sigma R+C ,
\end{equation}
where $\alpha$ is the Coulomb parameter, $\sigma$ the string tension,
and $C$ a constant. 

The correct distance between two points should be measured using geodesic 
distances. We take the distance $R$ between the source and sink to be equal 
to the index distance along the main axes of the skeleton. This seems a 
reasonable approximation in the well-defined phase with small curvature 
fluctuations. Using scalar field propagation \cite{HH2} one may calculate 
corrections which are expected to be small for our purposes.

To extract a reliable value for the string tension, we simulated the
system on an $8^{3}\times 4$ lattice. Our data rely on 30000 measurements 
after equilibration for the coupled system as well as for the pure gauge 
system on a flat simplicial lattice. 

\begin{figure}
\input{phasefig4a}
\vspace{1cm}
\input{phasefig4b}
\vspace{1cm}
\caption{\label{pots} Static quark potentials on an $8^{3}\times 4$
fluctuating lattice with $m_{p}^{2}=0.005$ (a) and for a flat simplicial
lattice with gravity switched off (b). The curves are fits to a confined
potential with periodicity taken into account. Error
bars arise from a jackknife procedure.}
\end{figure}
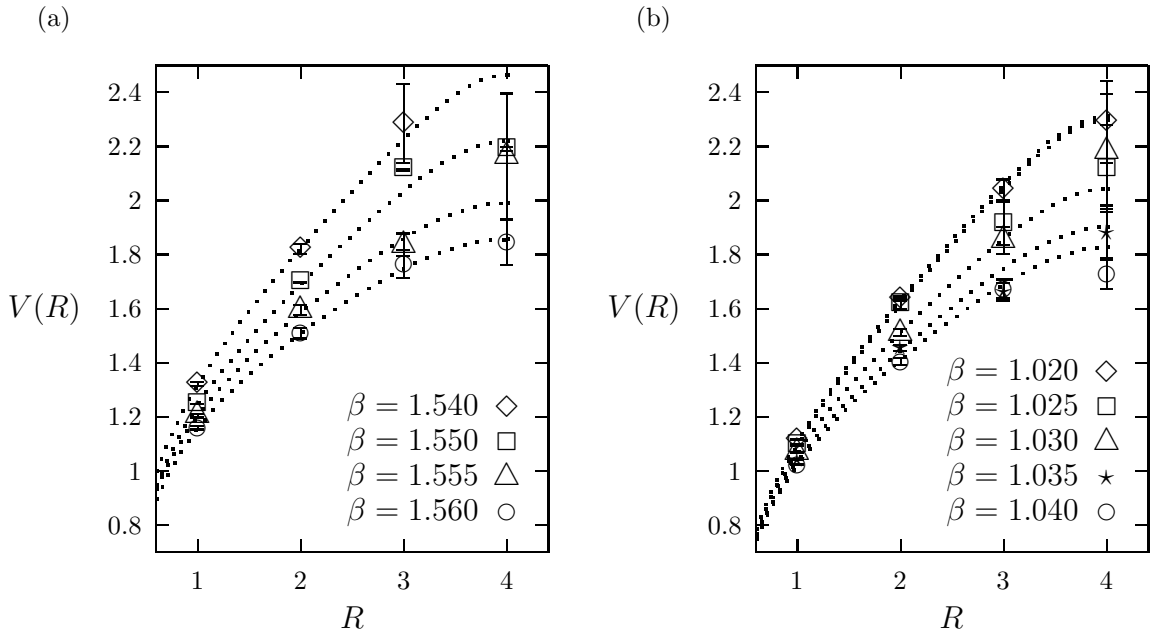

Figure~\ref{pots}(a) presents the data points for the confinement 
potentials for several gauge couplings in the presence of gravity
with $m_{p}^{2}=0.005$ in the ``entropy dominated'' phase, while 
Fig.~\ref{pots}(b)  depicts the situation with gravity switched off.
The dotted lines correspond to fits to the correlations in
Eq.~(\ref{eq:correl}) according to the potential of Eq.~(\ref{Vc})  
with the Coulomb parameter fixed to $\alpha=\frac{\pi}{12}$ and a 
mirror term included.

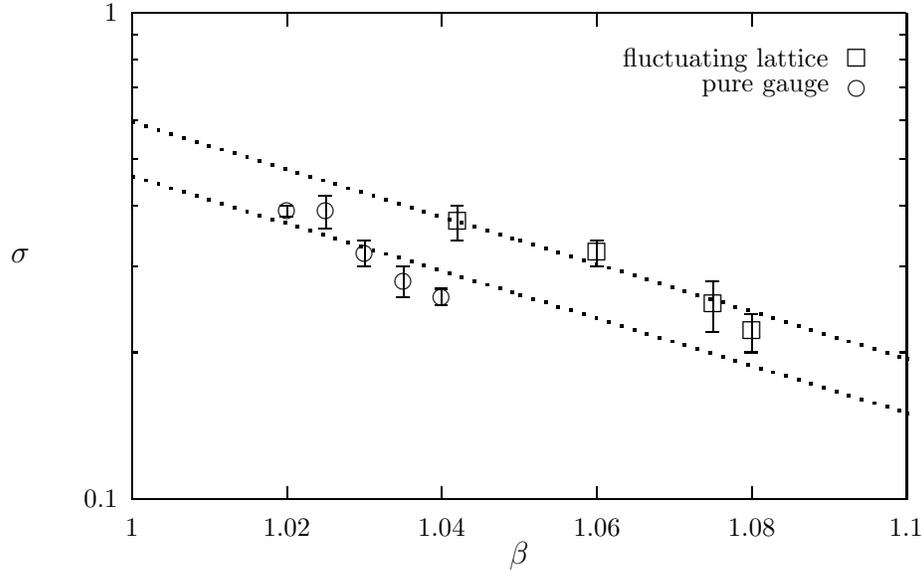
\begin{figure}
\input{phasefig5}
\vspace{1cm}
\caption{\label{scal} Scaling of the string tension $\sigma$
for the flat and the fluctuating lattice $(m_P^2 = 0.005)$.
$\sigma$ was extracted from fits to Eq.~(3.2) using $\alpha=\pi/12$.}
\end{figure}

Since we have extracted string tensions for several $\beta$ values,
we are in a position to study its scaling behavior. For pure gauge theory
$\beta$-functions are derived in the literature also for a simplicial
lattice \cite{DM}. We fit our string-tension data to the function 
\begin{equation} \label{betafunc}
\sigma_{\rm flat}=\frac{\sigma_{\rm phys}}{\Lambda^2_{\rm flat}}
\left(\frac{6\pi^2\sqrt{5}\beta}{11}\right)^{102/121}
\exp{\left(-\frac{6\pi^2\sqrt{5}\beta}{11}\right)} 
\end{equation}
and obtain a value for $\Lambda_{\rm flat}=0.0102(1)\sqrt{\sigma_{\rm phys}}$. 
This is in good agreement with an analysis of Wilson-loop ratios in the 
$T=0$ case, yielding 
\hbox{$\Lambda_{\rm flat}=0.008(1)\sqrt{\sigma_{\rm phys}}$}.
For the fluctuating lattice to our knowledge a $\beta$-function is not worked 
out, we are aware only of a recent study for the pure gravity case
within dynamical triangulation \cite{B}. Thus, we used as a starting point 
the above SU(2) function Eq.~(\ref{betafunc}). In Fig.~\ref{scal} we compare 
the string-tension scaling for the flat and the fluctuating lattices. 

To have both systems on the same scale, we rescaled the inverse gauge 
coupling \hbox{$\beta\to\langle W_t\rangle\beta$} of the fluctuating system. 
We find 
$\Lambda_{\rm grav} = 0.0090(1)\sqrt{\sigma_{\rm phys}}$ which is very 
similar to the flat case. Assuming that the pure-gauge $\beta$-function 
is a reasonable approximation for the full renormalization relation, our 
data show that the scaling window opens already at the $\beta$ values 
considered, similar to the pure SU(2) case \cite{DM}. As a consequence, the 
continuum limit could be performed along $\beta\to\infty$, eventually 
stopping at a finite value corresponding to a lattice spacing equal to 
the Planck length. The existence of a corridor to large $\beta$'s is 
indicated in the $(m_p, \beta)$ diagrams of Fig.~\ref{phadia}.

\section{Conclusion}
The boundary between the well-defined and the ill-defined phase of
Regge quantum gravity coupled to SU(2) gauge theory was studied with
the largest so far available statistics on $6^3\times 4$ lattices.
These lattices are already very CPU time intensive, because
the gravitational dynamics is very slow. Altogether, 
evidence was gained that the well-defined phase
is stable with the increase from $4^3\times 2$ to $6^3\times 4$.

Within the well-defined phase we find that the confinement mechanism from the 
non-Abelian gauge fields is not spoiled by quantum gravitational effects. 
This is not trivial, because it is was not clear how quantum gravity 
affects a gauge theory. In the confined phase we observed a potential 
linearly rising with $R$. Extracting string tension values for both the 
coupled system and for the pure gauge theory on a simplicial 
lattice without gravity we found a very similar scaling behavior. 
It indicates that gravity effects do not destroy the physics of
conventional asymptotically free field theories, even if the gravity-gauge 
coupling is large as in our situation. This gives hope that the physical 
limit can be approached through scaling from the investigated region
towards the Planck length. Additionally, to reproduce Newton's law one 
should demonstrate a diverging gravitational correlation length of
graviton propagators when approaching the boundary of the well-defined phase.

\newpage
\noindent

\newpage
\noindent

\newpage
\noindent

\newpage
\noindent

\newpage
\noindent

\newpage
\noindent
\begin{table}
\caption{\label{table1} Statistics for the $4^3\times 2$ lattices in units
 of 1k sweeps.}
\begin{tabular}{l|r||l|rrrrrrr}
 $m_p^2 \setminus \beta $ & 0.0 & $m_p^2 \setminus \beta $ 
 & 1.0 & 1.5 & 1.55 & 1.6 & 1.7 & 1.8  & 3.0     \\ \hline
$+0.0225 $ & 200  &
$+0.05   $ & 100 &     &      &     &      &     &          \\ \hline
$+0.02   $ & 200  & 
$+0.04   $ &  40  &     &      &     &      &     &          \\ \hline
$+0.0    $ & 200  &  
$+0.0350 $ & 167 &     &      &     &      &     &          \\ \hline
$-0.005  $ & 40   & 
$+0.03   $ & 200 & 100 & 100  &     &      &     &         \\ \hline
$-0.01   $ & 40   & 
$+0.025  $ & 100 & 200 &      & 140 &      &     & 200     \\ \hline
$-0.025  $ & 40   & 
$+0.0225 $ & 30   & 200 & 200  & 200 & 200  &     &         \\ \hline
$-0.03   $ & 40   &
$+0.005  $ &      &     &      &     &      &     & 100      \\ \hline
$-0.04   $ & 40   & 
$+0.0    $ &     & 35  &      &     &      &     &         \\  \hline
$-0.05   $ & 40   &
$-0.0025 $ &     & 200 &      & 200 &      &     &         \\ \hline
$-0.06   $ & 40   & 
$-0.005  $ & 60  & 200 &      & 200 &      & 100 & 200     \\ \hline
$-0.06375$ & 100  &   
$-0.0075 $ & 100 &     &      &     &      &     &         \\ \hline
$-0.06425$ & 200  &  
$-0.008  $ & 100 &     &      &     &      &     &         \\ \hline
$-0.065  $ &  18  &   
$-0.009  $ & 200 &     &      &     &      &     &         \\ \hline
$        $ &      &   
$-0.0105 $ & 100 &     &      &     &      &     &         \\
\end{tabular}
\end{table}

\newpage
\noindent
\begin{table}
\caption{\label{table2} Statistics for the $6^3\times 4$ lattices in units
 of 1k sweeps.}
\begin{tabular}{l|rrrrrrrrrrr} 
  $m_p^2 \setminus \beta$
          & 0.0 &0.073& 0.5& 0.8& 1.0& 1.2& 1.4& 1.55& 1.6& 1.65\\ 
\hline\tableline
$+0.03  $&  24 &     &    &    &152 &    &    &     & 88 & \\ \hline
$+0.0275$&     & 72  &    &    &    &    &    &     & 64 & \\ \hline
$+0.026 $&     &     &    &    & 48 &    &    &120  &208 & \\ \hline
$+0.025 $&     &     &    &    &    &    &    &136  &200 & \\ \hline
$+0.023 $&     &     &    &    &208 &    &    &144  &208 & \\ \hline
$+0.02  $&  56 &     &    &    &    &    &    &208  &    & \\ \hline
$+0.0175$&  23 &     &    &    &    &    &    &     &    & \\ \hline
$+0.015 $& 208 &     &    &    &    &    &    & 56  & 48 & \\ \hline
$+0.01  $&     &     &    &    &    &    &    &     &    & 56 \\ \hline
$-0.005 $&     &     &    &    &    &    &    &     &208 & \\ \hline
$-0.01  $&  24 &     &    & 40 &160 & 40 & 24 &     & 40 & \\ \hline
$-0.015 $&     &     &    &    &    &    &    &     & 32 & \\ \hline
$-0.02  $&  24 &     & 40 &    & 16 &    &    &     &    & \\ \hline
$-0.04  $&  40 &     &    &    &    &    &    &     &    & \\ \hline
$-0.05  $& 144 &     &    &    &    &    &    &     &    & \\ \hline
$-0.06  $&  40 &     &    &    &    &    &    &     &    & \\ 
\end{tabular}
\end{table}

\end{document}

%% file: phasefig3a.tex
\setlength{\unitlength}{0.240900pt}
\ifx\plotpoint\undefined\newsavebox{\plotpoint}\fi
\sbox{\plotpoint}{\rule[-0.200pt]{0.400pt}{0.400pt}}%
\begin{picture}(1500,900)(0,0)
\font\gnuplot=cmr10 at 10pt
\gnuplot
\sbox{\plotpoint}{\rule[-0.200pt]{0.400pt}{0.400pt}}%
\put(220.0,182.0){\rule[-0.200pt]{4.818pt}{0.400pt}}
\put(198,182){\makebox(0,0)[r]{-0.06}}
\put(1416.0,182.0){\rule[-0.200pt]{4.818pt}{0.400pt}}
\put(220.0,321.0){\rule[-0.200pt]{4.818pt}{0.400pt}}
\put(198,321){\makebox(0,0)[r]{-0.04}}
\put(1416.0,321.0){\rule[-0.200pt]{4.818pt}{0.400pt}}
\put(220.0,460.0){\rule[-0.200pt]{4.818pt}{0.400pt}}
\put(198,460){\makebox(0,0)[r]{-0.02}}
\put(1416.0,460.0){\rule[-0.200pt]{4.818pt}{0.400pt}}
\put(220.0,599.0){\rule[-0.200pt]{4.818pt}{0.400pt}}
\put(198,599){\makebox(0,0)[r]{0}}
\put(1416.0,599.0){\rule[-0.200pt]{4.818pt}{0.400pt}}
\put(220.0,738.0){\rule[-0.200pt]{4.818pt}{0.400pt}}
\put(198,738){\makebox(0,0)[r]{0.02}}
\put(1416.0,738.0){\rule[-0.200pt]{4.818pt}{0.400pt}}
\put(220.0,877.0){\rule[-0.200pt]{4.818pt}{0.400pt}}
\put(198,877){\makebox(0,0)[r]{0.04}}
\put(1416.0,877.0){\rule[-0.200pt]{4.818pt}{0.400pt}}
\put(288.0,113.0){\rule[-0.200pt]{0.400pt}{4.818pt}}
\put(288,68){\makebox(0,0){0}}
\put(288.0,857.0){\rule[-0.200pt]{0.400pt}{4.818pt}}
\put(423.0,113.0){\rule[-0.200pt]{0.400pt}{4.818pt}}
\put(423,68){\makebox(0,0){0.2}}
\put(423.0,857.0){\rule[-0.200pt]{0.400pt}{4.818pt}}
\put(558.0,113.0){\rule[-0.200pt]{0.400pt}{4.818pt}}
\put(558,68){\makebox(0,0){0.4}}
\put(558.0,857.0){\rule[-0.200pt]{0.400pt}{4.818pt}}
\put(693.0,113.0){\rule[-0.200pt]{0.400pt}{4.818pt}}
\put(693,68){\makebox(0,0){0.6}}
\put(693.0,857.0){\rule[-0.200pt]{0.400pt}{4.818pt}}
\put(828.0,113.0){\rule[-0.200pt]{0.400pt}{4.818pt}}
\put(828,68){\makebox(0,0){0.8}}
\put(828.0,857.0){\rule[-0.200pt]{0.400pt}{4.818pt}}
\put(963.0,113.0){\rule[-0.200pt]{0.400pt}{4.818pt}}
\put(963,68){\makebox(0,0){1}}
\put(963.0,857.0){\rule[-0.200pt]{0.400pt}{4.818pt}}
\put(1098.0,113.0){\rule[-0.200pt]{0.400pt}{4.818pt}}
\put(1098,68){\makebox(0,0){1.2}}
\put(1098.0,857.0){\rule[-0.200pt]{0.400pt}{4.818pt}}
\put(1233.0,113.0){\rule[-0.200pt]{0.400pt}{4.818pt}}
\put(1233,68){\makebox(0,0){1.4}}
\put(1233.0,857.0){\rule[-0.200pt]{0.400pt}{4.818pt}}
\put(1368.0,113.0){\rule[-0.200pt]{0.400pt}{4.818pt}}
\put(1368,68){\makebox(0,0){1.6}}
\put(1368.0,857.0){\rule[-0.200pt]{0.400pt}{4.818pt}}
\put(220.0,113.0){\rule[-0.200pt]{292.934pt}{0.400pt}}
\put(1436.0,113.0){\rule[-0.200pt]{0.400pt}{184.048pt}}
\put(220.0,877.0){\rule[-0.200pt]{292.934pt}{0.400pt}}
\put(60,980){\makebox(0,0){$(a)$}}
\put(45,599){\makebox(0,0){$m^2_p$}}
\put(828,10){\makebox(0,0){$\beta$}}
\put(558,669){\makebox(0,0)[l]{Well-defined}}
\put(558,828){\makebox(0,0)[l]{Ill-defined}}
\put(558,217){\makebox(0,0)[l]{Ill-defined}}
\put(220.0,113.0){\rule[-0.200pt]{0.400pt}{184.048pt}}
\sbox{\plotpoint}{\rule[-0.400pt]{0.800pt}{0.800pt}}%
\put(1301,252){\makebox(0,0)[r]{Stable}}
\put(1345,252){\raisebox{-.8pt}{\makebox(0,0){$\Diamond$}}}
\put(288,153){\raisebox{-.8pt}{\makebox(0,0){$\Diamond$}}}
\put(288,156){\raisebox{-.8pt}{\makebox(0,0){$\Diamond$}}}
\put(288,182){\raisebox{-.8pt}{\makebox(0,0){$\Diamond$}}}
\put(288,252){\raisebox{-.8pt}{\makebox(0,0){$\Diamond$}}}
\put(288,321){\raisebox{-.8pt}{\makebox(0,0){$\Diamond$}}}
\put(288,391){\raisebox{-.8pt}{\makebox(0,0){$\Diamond$}}}
\put(288,426){\raisebox{-.8pt}{\makebox(0,0){$\Diamond$}}}
\put(288,530){\raisebox{-.8pt}{\makebox(0,0){$\Diamond$}}}
\put(288,564){\raisebox{-.8pt}{\makebox(0,0){$\Diamond$}}}
\put(288,599){\raisebox{-.8pt}{\makebox(0,0){$\Diamond$}}}
\put(288,738){\raisebox{-.8pt}{\makebox(0,0){$\Diamond$}}}
\put(963,537){\raisebox{-.8pt}{\makebox(0,0){$\Diamond$}}}
\put(963,544){\raisebox{-.8pt}{\makebox(0,0){$\Diamond$}}}
\put(963,547){\raisebox{-.8pt}{\makebox(0,0){$\Diamond$}}}
\put(963,564){\raisebox{-.8pt}{\makebox(0,0){$\Diamond$}}}
\put(963,755){\raisebox{-.8pt}{\makebox(0,0){$\Diamond$}}}
\put(963,773){\raisebox{-.8pt}{\makebox(0,0){$\Diamond$}}}
\put(963,808){\raisebox{-.8pt}{\makebox(0,0){$\Diamond$}}}
\put(1301,582){\raisebox{-.8pt}{\makebox(0,0){$\Diamond$}}}
\put(1301,599){\raisebox{-.8pt}{\makebox(0,0){$\Diamond$}}}
\put(1301,755){\raisebox{-.8pt}{\makebox(0,0){$\Diamond$}}}
\put(1335,755){\raisebox{-.8pt}{\makebox(0,0){$\Diamond$}}}
\put(1368,582){\raisebox{-.8pt}{\makebox(0,0){$\Diamond$}}}
\put(1368,755){\raisebox{-.8pt}{\makebox(0,0){$\Diamond$}}}
\sbox{\plotpoint}{\rule[-0.500pt]{1.000pt}{1.000pt}}%
\put(1301,207){\makebox(0,0)[r]{Unstable}}
\put(1345,207){\makebox(0,0){$+$}}
\put(288,148){\makebox(0,0){$+$}}
\put(288,755){\makebox(0,0){$+$}}
\put(963,526){\makebox(0,0){$+$}}
\put(963,842){\makebox(0,0){$+$}}
\put(1301,564){\makebox(0,0){$+$}}
\put(1301,773){\makebox(0,0){$+$}}
\put(1301,808){\makebox(0,0){$+$}}
\put(1335,808){\makebox(0,0){$+$}}
\put(1368,564){\makebox(0,0){$+$}}
\put(1368,773){\makebox(0,0){$+$}}
\put(288,150){\usebox{\plotpoint}}
\multiput(288,150)(18.075,10.202){38}{\usebox{\plotpoint}}
\multiput(963,531)(20.597,2.559){16}{\usebox{\plotpoint}}
\multiput(1301,573)(20.756,0.000){3}{\usebox{\plotpoint}}
\multiput(1368,573)(20.756,0.000){4}{\usebox{\plotpoint}}
\put(1436,573){\usebox{\plotpoint}}
\multiput(1436,768)(-20.720,-1.219){4}{\usebox{\plotpoint}}
\multiput(1368,764)(-20.756,0.000){3}{\usebox{\plotpoint}}
\multiput(1301,764)(-20.426,3.686){17}{\usebox{\plotpoint}}
\multiput(963,825)(-20.618,-2.383){32}{\usebox{\plotpoint}}
\put(288,747){\usebox{\plotpoint}}
\sbox{\plotpoint}{\rule[-0.400pt]{0.800pt}{0.800pt}}%
\put(1318,842){\usebox{\plotpoint}}
\put(1318.0,460.0){\rule[-0.400pt]{0.800pt}{92.024pt}}
\sbox{\plotpoint}{\rule[-0.200pt]{0.400pt}{0.400pt}}%
\put(288,113){\usebox{\plotpoint}}
\multiput(288,113)(0.000,20.756){37}{\usebox{\plotpoint}}
\put(288,877){\usebox{\plotpoint}}
\put(220,599){\usebox{\plotpoint}}
\put(220.00,599.00){\usebox{\plotpoint}}
\put(240.76,599.00){\usebox{\plotpoint}}
\multiput(245,599)(20.756,0.000){0}{\usebox{\plotpoint}}
\put(261.51,599.00){\usebox{\plotpoint}}
\multiput(269,599)(20.756,0.000){0}{\usebox{\plotpoint}}
\put(282.27,599.00){\usebox{\plotpoint}}
\put(303.02,599.00){\usebox{\plotpoint}}
\multiput(306,599)(20.756,0.000){0}{\usebox{\plotpoint}}
\put(323.78,599.00){\usebox{\plotpoint}}
\multiput(331,599)(20.756,0.000){0}{\usebox{\plotpoint}}
\put(344.53,599.00){\usebox{\plotpoint}}
\put(365.29,599.00){\usebox{\plotpoint}}
\multiput(367,599)(20.756,0.000){0}{\usebox{\plotpoint}}
\put(386.04,599.00){\usebox{\plotpoint}}
\multiput(392,599)(20.756,0.000){0}{\usebox{\plotpoint}}
\put(406.80,599.00){\usebox{\plotpoint}}
\put(427.55,599.00){\usebox{\plotpoint}}
\multiput(429,599)(20.756,0.000){0}{\usebox{\plotpoint}}
\put(448.31,599.00){\usebox{\plotpoint}}
\multiput(453,599)(20.756,0.000){0}{\usebox{\plotpoint}}
\put(469.07,599.00){\usebox{\plotpoint}}
\put(489.82,599.00){\usebox{\plotpoint}}
\multiput(490,599)(20.756,0.000){0}{\usebox{\plotpoint}}
\put(510.58,599.00){\usebox{\plotpoint}}
\multiput(515,599)(20.756,0.000){0}{\usebox{\plotpoint}}
\put(531.33,599.00){\usebox{\plotpoint}}
\multiput(539,599)(20.756,0.000){0}{\usebox{\plotpoint}}
\put(552.09,599.00){\usebox{\plotpoint}}
\put(572.84,599.00){\usebox{\plotpoint}}
\multiput(576,599)(20.756,0.000){0}{\usebox{\plotpoint}}
\put(593.60,599.00){\usebox{\plotpoint}}
\multiput(601,599)(20.756,0.000){0}{\usebox{\plotpoint}}
\put(614.35,599.00){\usebox{\plotpoint}}
\put(635.11,599.00){\usebox{\plotpoint}}
\multiput(638,599)(20.756,0.000){0}{\usebox{\plotpoint}}
\put(655.87,599.00){\usebox{\plotpoint}}
\multiput(662,599)(20.756,0.000){0}{\usebox{\plotpoint}}
\put(676.62,599.00){\usebox{\plotpoint}}
\put(697.38,599.00){\usebox{\plotpoint}}
\multiput(699,599)(20.756,0.000){0}{\usebox{\plotpoint}}
\put(718.13,599.00){\usebox{\plotpoint}}
\multiput(724,599)(20.756,0.000){0}{\usebox{\plotpoint}}
\put(738.89,599.00){\usebox{\plotpoint}}
\put(759.64,599.00){\usebox{\plotpoint}}
\multiput(760,599)(20.756,0.000){0}{\usebox{\plotpoint}}
\put(780.40,599.00){\usebox{\plotpoint}}
\multiput(785,599)(20.756,0.000){0}{\usebox{\plotpoint}}
\put(801.15,599.00){\usebox{\plotpoint}}
\put(821.91,599.00){\usebox{\plotpoint}}
\multiput(822,599)(20.756,0.000){0}{\usebox{\plotpoint}}
\put(842.66,599.00){\usebox{\plotpoint}}
\multiput(846,599)(20.756,0.000){0}{\usebox{\plotpoint}}
\put(863.42,599.00){\usebox{\plotpoint}}
\multiput(871,599)(20.756,0.000){0}{\usebox{\plotpoint}}
\put(884.18,599.00){\usebox{\plotpoint}}
\put(904.93,599.00){\usebox{\plotpoint}}
\multiput(908,599)(20.756,0.000){0}{\usebox{\plotpoint}}
\put(925.69,599.00){\usebox{\plotpoint}}
\multiput(932,599)(20.756,0.000){0}{\usebox{\plotpoint}}
\put(946.44,599.00){\usebox{\plotpoint}}
\put(967.20,599.00){\usebox{\plotpoint}}
\multiput(969,599)(20.756,0.000){0}{\usebox{\plotpoint}}
\put(987.95,599.00){\usebox{\plotpoint}}
\multiput(994,599)(20.756,0.000){0}{\usebox{\plotpoint}}
\put(1008.71,599.00){\usebox{\plotpoint}}
\put(1029.46,599.00){\usebox{\plotpoint}}
\multiput(1031,599)(20.756,0.000){0}{\usebox{\plotpoint}}
\put(1050.22,599.00){\usebox{\plotpoint}}
\multiput(1055,599)(20.756,0.000){0}{\usebox{\plotpoint}}
\put(1070.98,599.00){\usebox{\plotpoint}}
\put(1091.73,599.00){\usebox{\plotpoint}}
\multiput(1092,599)(20.756,0.000){0}{\usebox{\plotpoint}}
\put(1112.49,599.00){\usebox{\plotpoint}}
\multiput(1117,599)(20.756,0.000){0}{\usebox{\plotpoint}}
\put(1133.24,599.00){\usebox{\plotpoint}}
\multiput(1141,599)(20.756,0.000){0}{\usebox{\plotpoint}}
\put(1154.00,599.00){\usebox{\plotpoint}}
\put(1174.75,599.00){\usebox{\plotpoint}}
\multiput(1178,599)(20.756,0.000){0}{\usebox{\plotpoint}}
\put(1195.51,599.00){\usebox{\plotpoint}}
\multiput(1203,599)(20.756,0.000){0}{\usebox{\plotpoint}}
\put(1216.26,599.00){\usebox{\plotpoint}}
\put(1237.02,599.00){\usebox{\plotpoint}}
\multiput(1239,599)(20.756,0.000){0}{\usebox{\plotpoint}}
\put(1257.77,599.00){\usebox{\plotpoint}}
\multiput(1264,599)(20.756,0.000){0}{\usebox{\plotpoint}}
\put(1278.53,599.00){\usebox{\plotpoint}}
\put(1299.29,599.00){\usebox{\plotpoint}}
\multiput(1301,599)(20.756,0.000){0}{\usebox{\plotpoint}}
\put(1320.04,599.00){\usebox{\plotpoint}}
\multiput(1325,599)(20.756,0.000){0}{\usebox{\plotpoint}}
\put(1340.80,599.00){\usebox{\plotpoint}}
\put(1361.55,599.00){\usebox{\plotpoint}}
\multiput(1362,599)(20.756,0.000){0}{\usebox{\plotpoint}}
\put(1382.31,599.00){\usebox{\plotpoint}}
\multiput(1387,599)(20.756,0.000){0}{\usebox{\plotpoint}}
\put(1403.06,599.00){\usebox{\plotpoint}}
\put(1423.82,599.00){\usebox{\plotpoint}}
\multiput(1424,599)(20.756,0.000){0}{\usebox{\plotpoint}}
\put(1436,599){\usebox{\plotpoint}}
\end{picture}

%% file: phasefig3b.tex
\setlength{\unitlength}{0.240900pt}
\ifx\plotpoint\undefined\newsavebox{\plotpoint}\fi
\sbox{\plotpoint}{\rule[-0.200pt]{0.400pt}{0.400pt}}%
\begin{picture}(1500,900)(0,0)
\font\gnuplot=cmr10 at 10pt
\gnuplot
\sbox{\plotpoint}{\rule[-0.200pt]{0.400pt}{0.400pt}}%
\put(220.0,182.0){\rule[-0.200pt]{4.818pt}{0.400pt}}
\put(198,182){\makebox(0,0)[r]{-0.06}}
\put(1416.0,182.0){\rule[-0.200pt]{4.818pt}{0.400pt}}
\put(220.0,321.0){\rule[-0.200pt]{4.818pt}{0.400pt}}
\put(198,321){\makebox(0,0)[r]{-0.04}}
\put(1416.0,321.0){\rule[-0.200pt]{4.818pt}{0.400pt}}
\put(220.0,460.0){\rule[-0.200pt]{4.818pt}{0.400pt}}
\put(198,460){\makebox(0,0)[r]{-0.02}}
\put(1416.0,460.0){\rule[-0.200pt]{4.818pt}{0.400pt}}
\put(220.0,599.0){\rule[-0.200pt]{4.818pt}{0.400pt}}
\put(198,599){\makebox(0,0)[r]{0}}
\put(1416.0,599.0){\rule[-0.200pt]{4.818pt}{0.400pt}}
\put(220.0,738.0){\rule[-0.200pt]{4.818pt}{0.400pt}}
\put(198,738){\makebox(0,0)[r]{0.02}}
\put(1416.0,738.0){\rule[-0.200pt]{4.818pt}{0.400pt}}
\put(220.0,877.0){\rule[-0.200pt]{4.818pt}{0.400pt}}
\put(198,877){\makebox(0,0)[r]{0.04}}
\put(1416.0,877.0){\rule[-0.200pt]{4.818pt}{0.400pt}}
\put(288.0,113.0){\rule[-0.200pt]{0.400pt}{4.818pt}}
\put(288,68){\makebox(0,0){0}}
\put(288.0,857.0){\rule[-0.200pt]{0.400pt}{4.818pt}}
\put(423.0,113.0){\rule[-0.200pt]{0.400pt}{4.818pt}}
\put(423,68){\makebox(0,0){0.2}}
\put(423.0,857.0){\rule[-0.200pt]{0.400pt}{4.818pt}}
\put(558.0,113.0){\rule[-0.200pt]{0.400pt}{4.818pt}}
\put(558,68){\makebox(0,0){0.4}}
\put(558.0,857.0){\rule[-0.200pt]{0.400pt}{4.818pt}}
\put(693.0,113.0){\rule[-0.200pt]{0.400pt}{4.818pt}}
\put(693,68){\makebox(0,0){0.6}}
\put(693.0,857.0){\rule[-0.200pt]{0.400pt}{4.818pt}}
\put(828.0,113.0){\rule[-0.200pt]{0.400pt}{4.818pt}}
\put(828,68){\makebox(0,0){0.8}}
\put(828.0,857.0){\rule[-0.200pt]{0.400pt}{4.818pt}}
\put(963.0,113.0){\rule[-0.200pt]{0.400pt}{4.818pt}}
\put(963,68){\makebox(0,0){1}}
\put(963.0,857.0){\rule[-0.200pt]{0.400pt}{4.818pt}}
\put(1098.0,113.0){\rule[-0.200pt]{0.400pt}{4.818pt}}
\put(1098,68){\makebox(0,0){1.2}}
\put(1098.0,857.0){\rule[-0.200pt]{0.400pt}{4.818pt}}
\put(1233.0,113.0){\rule[-0.200pt]{0.400pt}{4.818pt}}
\put(1233,68){\makebox(0,0){1.4}}
\put(1233.0,857.0){\rule[-0.200pt]{0.400pt}{4.818pt}}
\put(1368.0,113.0){\rule[-0.200pt]{0.400pt}{4.818pt}}
\put(1368,68){\makebox(0,0){1.6}}
\put(1368.0,857.0){\rule[-0.200pt]{0.400pt}{4.818pt}}
\put(220.0,113.0){\rule[-0.200pt]{292.934pt}{0.400pt}}
\put(1436.0,113.0){\rule[-0.200pt]{0.400pt}{184.048pt}}
\put(220.0,877.0){\rule[-0.200pt]{292.934pt}{0.400pt}}
\put(60,980){\makebox(0,0){$(b)$}}
\put(60,599){\makebox(0,0){$m^2_p$}}
\put(828,10){\makebox(0,0){$\beta$}}
\put(558,669){\makebox(0,0)[l]{Well-defined}}
\put(558,808){\makebox(0,0)[l]{Ill-defined}}
\put(558,217){\makebox(0,0)[l]{Ill-defined}}
\put(220.0,113.0){\rule[-0.200pt]{0.400pt}{184.048pt}}
\sbox{\plotpoint}{\rule[-0.400pt]{0.800pt}{0.800pt}}%
\put(1301,252){\makebox(0,0)[r]{Stable}}
\put(1345,252){\raisebox{-.8pt}{\makebox(0,0){$\Diamond$}}}
\put(288,703){\raisebox{-.8pt}{\makebox(0,0){$\Diamond$}}}
\put(288,530){\raisebox{-.8pt}{\makebox(0,0){$\Diamond$}}}
\put(288,460){\raisebox{-.8pt}{\makebox(0,0){$\Diamond$}}}
\put(288,321){\raisebox{-.8pt}{\makebox(0,0){$\Diamond$}}}
\put(288,252){\raisebox{-.8pt}{\makebox(0,0){$\Diamond$}}}
\put(625,460){\raisebox{-.8pt}{\makebox(0,0){$\Diamond$}}}
\put(828,530){\raisebox{-.8pt}{\makebox(0,0){$\Diamond$}}}
\put(963,759){\raisebox{-.8pt}{\makebox(0,0){$\Diamond$}}}
\put(963,530){\raisebox{-.8pt}{\makebox(0,0){$\Diamond$}}}
\put(1098,530){\raisebox{-.8pt}{\makebox(0,0){$\Diamond$}}}
\put(1233,530){\raisebox{-.8pt}{\makebox(0,0){$\Diamond$}}}
\put(1335,703){\raisebox{-.8pt}{\makebox(0,0){$\Diamond$}}}
\put(1335,738){\raisebox{-.8pt}{\makebox(0,0){$\Diamond$}}}
\put(1368,773){\raisebox{-.8pt}{\makebox(0,0){$\Diamond$}}}
\put(1368,759){\raisebox{-.8pt}{\makebox(0,0){$\Diamond$}}}
\put(1368,703){\raisebox{-.8pt}{\makebox(0,0){$\Diamond$}}}
\put(1368,564){\raisebox{-.8pt}{\makebox(0,0){$\Diamond$}}}
\put(1402,669){\raisebox{-.8pt}{\makebox(0,0){$\Diamond$}}}
\sbox{\plotpoint}{\rule[-0.500pt]{1.000pt}{1.000pt}}%
\put(1301,207){\makebox(0,0)[r]{Unstable}}
\put(1345,207){\makebox(0,0){$+$}}
\put(288,808){\makebox(0,0){$+$}}
\put(288,738){\makebox(0,0){$+$}}
\put(288,721){\makebox(0,0){$+$}}
\put(288,182){\makebox(0,0){$+$}}
\put(337,790){\makebox(0,0){$+$}}
\put(963,808){\makebox(0,0){$+$}}
\put(963,780){\makebox(0,0){$+$}}
\put(963,460){\makebox(0,0){$+$}}
\put(1335,780){\makebox(0,0){$+$}}
\put(1335,773){\makebox(0,0){$+$}}
\put(1335,759){\makebox(0,0){$+$}}
\put(1368,808){\makebox(0,0){$+$}}
\put(1368,790){\makebox(0,0){$+$}}
\put(1368,780){\makebox(0,0){$+$}}
\put(1368,530){\makebox(0,0){$+$}}
\put(1368,495){\makebox(0,0){$+$}}
\put(288,712){\usebox{\plotpoint}}
\multiput(288,712)(20.682,1.746){33}{\usebox{\plotpoint}}
\multiput(963,769)(20.734,-0.948){18}{\usebox{\plotpoint}}
\multiput(1335,752)(17.511,11.143){2}{\usebox{\plotpoint}}
\multiput(1368,773)(20.756,0.000){3}{\usebox{\plotpoint}}
\put(1436,773){\usebox{\plotpoint}}
\multiput(1436,556)(-20.576,-2.723){4}{\usebox{\plotpoint}}
\multiput(1368,547)(-20.587,-2.643){19}{\usebox{\plotpoint}}
\multiput(963,495)(-19.192,-7.904){36}{\usebox{\plotpoint}}
\put(288,217){\usebox{\plotpoint}}
\sbox{\plotpoint}{\rule[-0.400pt]{0.800pt}{0.800pt}}%
\put(1352,842){\usebox{\plotpoint}}
\put(1352.0,460.0){\rule[-0.400pt]{0.800pt}{92.024pt}}
\sbox{\plotpoint}{\rule[-0.200pt]{0.400pt}{0.400pt}}%
\put(288,113){\usebox{\plotpoint}}
\multiput(288,113)(0.000,20.756){37}{\usebox{\plotpoint}}
\put(288,877){\usebox{\plotpoint}}
\put(220,599){\usebox{\plotpoint}}
\put(220.00,599.00){\usebox{\plotpoint}}
\put(240.76,599.00){\usebox{\plotpoint}}
\multiput(245,599)(20.756,0.000){0}{\usebox{\plotpoint}}
\put(261.51,599.00){\usebox{\plotpoint}}
\multiput(269,599)(20.756,0.000){0}{\usebox{\plotpoint}}
\put(282.27,599.00){\usebox{\plotpoint}}
\put(303.02,599.00){\usebox{\plotpoint}}
\multiput(306,599)(20.756,0.000){0}{\usebox{\plotpoint}}
\put(323.78,599.00){\usebox{\plotpoint}}
\multiput(331,599)(20.756,0.000){0}{\usebox{\plotpoint}}
\put(344.53,599.00){\usebox{\plotpoint}}
\put(365.29,599.00){\usebox{\plotpoint}}
\multiput(367,599)(20.756,0.000){0}{\usebox{\plotpoint}}
\put(386.04,599.00){\usebox{\plotpoint}}
\multiput(392,599)(20.756,0.000){0}{\usebox{\plotpoint}}
\put(406.80,599.00){\usebox{\plotpoint}}
\put(427.55,599.00){\usebox{\plotpoint}}
\multiput(429,599)(20.756,0.000){0}{\usebox{\plotpoint}}
\put(448.31,599.00){\usebox{\plotpoint}}
\multiput(453,599)(20.756,0.000){0}{\usebox{\plotpoint}}
\put(469.07,599.00){\usebox{\plotpoint}}
\put(489.82,599.00){\usebox{\plotpoint}}
\multiput(490,599)(20.756,0.000){0}{\usebox{\plotpoint}}
\put(510.58,599.00){\usebox{\plotpoint}}
\multiput(515,599)(20.756,0.000){0}{\usebox{\plotpoint}}
\put(531.33,599.00){\usebox{\plotpoint}}
\multiput(539,599)(20.756,0.000){0}{\usebox{\plotpoint}}
\put(552.09,599.00){\usebox{\plotpoint}}
\put(572.84,599.00){\usebox{\plotpoint}}
\multiput(576,599)(20.756,0.000){0}{\usebox{\plotpoint}}
\put(593.60,599.00){\usebox{\plotpoint}}
\multiput(601,599)(20.756,0.000){0}{\usebox{\plotpoint}}
\put(614.35,599.00){\usebox{\plotpoint}}
\put(635.11,599.00){\usebox{\plotpoint}}
\multiput(638,599)(20.756,0.000){0}{\usebox{\plotpoint}}
\put(655.87,599.00){\usebox{\plotpoint}}
\multiput(662,599)(20.756,0.000){0}{\usebox{\plotpoint}}
\put(676.62,599.00){\usebox{\plotpoint}}
\put(697.38,599.00){\usebox{\plotpoint}}
\multiput(699,599)(20.756,0.000){0}{\usebox{\plotpoint}}
\put(718.13,599.00){\usebox{\plotpoint}}
\multiput(724,599)(20.756,0.000){0}{\usebox{\plotpoint}}
\put(738.89,599.00){\usebox{\plotpoint}}
\put(759.64,599.00){\usebox{\plotpoint}}
\multiput(760,599)(20.756,0.000){0}{\usebox{\plotpoint}}
\put(780.40,599.00){\usebox{\plotpoint}}
\multiput(785,599)(20.756,0.000){0}{\usebox{\plotpoint}}
\put(801.15,599.00){\usebox{\plotpoint}}
\put(821.91,599.00){\usebox{\plotpoint}}
\multiput(822,599)(20.756,0.000){0}{\usebox{\plotpoint}}
\put(842.66,599.00){\usebox{\plotpoint}}
\multiput(846,599)(20.756,0.000){0}{\usebox{\plotpoint}}
\put(863.42,599.00){\usebox{\plotpoint}}
\multiput(871,599)(20.756,0.000){0}{\usebox{\plotpoint}}
\put(884.18,599.00){\usebox{\plotpoint}}
\put(904.93,599.00){\usebox{\plotpoint}}
\multiput(908,599)(20.756,0.000){0}{\usebox{\plotpoint}}
\put(925.69,599.00){\usebox{\plotpoint}}
\multiput(932,599)(20.756,0.000){0}{\usebox{\plotpoint}}
\put(946.44,599.00){\usebox{\plotpoint}}
\put(967.20,599.00){\usebox{\plotpoint}}
\multiput(969,599)(20.756,0.000){0}{\usebox{\plotpoint}}
\put(987.95,599.00){\usebox{\plotpoint}}
\multiput(994,599)(20.756,0.000){0}{\usebox{\plotpoint}}
\put(1008.71,599.00){\usebox{\plotpoint}}
\put(1029.46,599.00){\usebox{\plotpoint}}
\multiput(1031,599)(20.756,0.000){0}{\usebox{\plotpoint}}
\put(1050.22,599.00){\usebox{\plotpoint}}
\multiput(1055,599)(20.756,0.000){0}{\usebox{\plotpoint}}
\put(1070.98,599.00){\usebox{\plotpoint}}
\put(1091.73,599.00){\usebox{\plotpoint}}
\multiput(1092,599)(20.756,0.000){0}{\usebox{\plotpoint}}
\put(1112.49,599.00){\usebox{\plotpoint}}
\multiput(1117,599)(20.756,0.000){0}{\usebox{\plotpoint}}
\put(1133.24,599.00){\usebox{\plotpoint}}
\multiput(1141,599)(20.756,0.000){0}{\usebox{\plotpoint}}
\put(1154.00,599.00){\usebox{\plotpoint}}
\put(1174.75,599.00){\usebox{\plotpoint}}
\multiput(1178,599)(20.756,0.000){0}{\usebox{\plotpoint}}
\put(1195.51,599.00){\usebox{\plotpoint}}
\multiput(1203,599)(20.756,0.000){0}{\usebox{\plotpoint}}
\put(1216.26,599.00){\usebox{\plotpoint}}
\put(1237.02,599.00){\usebox{\plotpoint}}
\multiput(1239,599)(20.756,0.000){0}{\usebox{\plotpoint}}
\put(1257.77,599.00){\usebox{\plotpoint}}
\multiput(1264,599)(20.756,0.000){0}{\usebox{\plotpoint}}
\put(1278.53,599.00){\usebox{\plotpoint}}
\put(1299.29,599.00){\usebox{\plotpoint}}
\multiput(1301,599)(20.756,0.000){0}{\usebox{\plotpoint}}
\put(1320.04,599.00){\usebox{\plotpoint}}
\multiput(1325,599)(20.756,0.000){0}{\usebox{\plotpoint}}
\put(1340.80,599.00){\usebox{\plotpoint}}
\put(1361.55,599.00){\usebox{\plotpoint}}
\multiput(1362,599)(20.756,0.000){0}{\usebox{\plotpoint}}
\put(1382.31,599.00){\usebox{\plotpoint}}
\multiput(1387,599)(20.756,0.000){0}{\usebox{\plotpoint}}
\put(1403.06,599.00){\usebox{\plotpoint}}
\put(1423.82,599.00){\usebox{\plotpoint}}
\multiput(1424,599)(20.756,0.000){0}{\usebox{\plotpoint}}
\put(1436,599){\usebox{\plotpoint}}
\end{picture}

%% file: phasefig4a.tex
\setlength{\unitlength}{0.240900pt}
\ifx\plotpoint\undefined\newsavebox{\plotpoint}\fi
\sbox{\plotpoint}{\rule[-0.200pt]{0.400pt}{0.400pt}}%
\begin{picture}(900,900)(0,0)
\font\gnuplot=cmr10 at 10pt
\gnuplot
\sbox{\plotpoint}{\rule[-0.200pt]{0.400pt}{0.400pt}}%
\put(220.0,155.0){\rule[-0.200pt]{4.818pt}{0.400pt}}
\put(198,155){\makebox(0,0)[r]{0.8}}
\put(816.0,155.0){\rule[-0.200pt]{4.818pt}{0.400pt}}
\put(220.0,240.0){\rule[-0.200pt]{4.818pt}{0.400pt}}
\put(198,240){\makebox(0,0)[r]{1}}
\put(816.0,240.0){\rule[-0.200pt]{4.818pt}{0.400pt}}
\put(220.0,325.0){\rule[-0.200pt]{4.818pt}{0.400pt}}
\put(198,325){\makebox(0,0)[r]{1.2}}
\put(816.0,325.0){\rule[-0.200pt]{4.818pt}{0.400pt}}
\put(220.0,410.0){\rule[-0.200pt]{4.818pt}{0.400pt}}
\put(198,410){\makebox(0,0)[r]{1.4}}
\put(816.0,410.0){\rule[-0.200pt]{4.818pt}{0.400pt}}
\put(220.0,495.0){\rule[-0.200pt]{4.818pt}{0.400pt}}
\put(198,495){\makebox(0,0)[r]{1.6}}
\put(816.0,495.0){\rule[-0.200pt]{4.818pt}{0.400pt}}
\put(220.0,580.0){\rule[-0.200pt]{4.818pt}{0.400pt}}
\put(198,580){\makebox(0,0)[r]{1.8}}
\put(816.0,580.0){\rule[-0.200pt]{4.818pt}{0.400pt}}
\put(220.0,665.0){\rule[-0.200pt]{4.818pt}{0.400pt}}
\put(198,665){\makebox(0,0)[r]{2}}
\put(816.0,665.0){\rule[-0.200pt]{4.818pt}{0.400pt}}
\put(220.0,750.0){\rule[-0.200pt]{4.818pt}{0.400pt}}
\put(198,750){\makebox(0,0)[r]{2.2}}
\put(816.0,750.0){\rule[-0.200pt]{4.818pt}{0.400pt}}
\put(220.0,835.0){\rule[-0.200pt]{4.818pt}{0.400pt}}
\put(198,835){\makebox(0,0)[r]{2.4}}
\put(816.0,835.0){\rule[-0.200pt]{4.818pt}{0.400pt}}
\put(285.0,113.0){\rule[-0.200pt]{0.400pt}{4.818pt}}
\put(285,68){\makebox(0,0){1}}
\put(285.0,857.0){\rule[-0.200pt]{0.400pt}{4.818pt}}
\put(447.0,113.0){\rule[-0.200pt]{0.400pt}{4.818pt}}
\put(447,68){\makebox(0,0){2}}
\put(447.0,857.0){\rule[-0.200pt]{0.400pt}{4.818pt}}
\put(609.0,113.0){\rule[-0.200pt]{0.400pt}{4.818pt}}
\put(609,68){\makebox(0,0){3}}
\put(609.0,857.0){\rule[-0.200pt]{0.400pt}{4.818pt}}
\put(771.0,113.0){\rule[-0.200pt]{0.400pt}{4.818pt}}
\put(771,68){\makebox(0,0){4}}
\put(771.0,857.0){\rule[-0.200pt]{0.400pt}{4.818pt}}
\put(220.0,113.0){\rule[-0.200pt]{148.394pt}{0.400pt}}
\put(836.0,113.0){\rule[-0.200pt]{0.400pt}{184.048pt}}
\put(220.0,877.0){\rule[-0.200pt]{148.394pt}{0.400pt}}
\put(60,950){\makebox(0,0){(a)}}
\put(45,495){\makebox(0,0){$V(R)$}}
\put(528,10){\makebox(0,0){$R$}}
\put(220.0,113.0){\rule[-0.200pt]{0.400pt}{184.048pt}}
\put(727,340){\makebox(0,0)[r]{$\beta=1.540$}}
\put(771,340){\raisebox{-.8pt}{\makebox(0,0){$\Diamond$}}}
\put(285,377){\raisebox{-.8pt}{\makebox(0,0){$\Diamond$}}}
\put(447,589){\raisebox{-.8pt}{\makebox(0,0){$\Diamond$}}}
\put(609,786){\raisebox{-.8pt}{\makebox(0,0){$\Diamond$}}}
\put(285.0,374.0){\rule[-0.200pt]{0.400pt}{1.445pt}}
\put(275.0,374.0){\rule[-0.200pt]{4.818pt}{0.400pt}}
\put(275.0,380.0){\rule[-0.200pt]{4.818pt}{0.400pt}}
\put(447.0,581.0){\rule[-0.200pt]{0.400pt}{3.854pt}}
\put(437.0,581.0){\rule[-0.200pt]{4.818pt}{0.400pt}}
\put(437.0,597.0){\rule[-0.200pt]{4.818pt}{0.400pt}}
\put(609.0,724.0){\rule[-0.200pt]{0.400pt}{29.872pt}}
\put(599.0,724.0){\rule[-0.200pt]{4.818pt}{0.400pt}}
\put(599.0,848.0){\rule[-0.200pt]{4.818pt}{0.400pt}}
\put(727,285){\makebox(0,0)[r]{$\beta=1.550$}}
\put(771,285){\raisebox{-.8pt}{\makebox(0,0){$\Box$}}}
\put(285,345){\raisebox{-.8pt}{\makebox(0,0){$\Box$}}}
\put(447,537){\raisebox{-.8pt}{\makebox(0,0){$\Box$}}}
\put(609,714){\raisebox{-.8pt}{\makebox(0,0){$\Box$}}}
\put(771,746){\raisebox{-.8pt}{\makebox(0,0){$\Box$}}}
\put(285,345){\usebox{\plotpoint}}
\put(275.0,345.0){\rule[-0.200pt]{4.818pt}{0.400pt}}
\put(275.0,345.0){\rule[-0.200pt]{4.818pt}{0.400pt}}
\put(447.0,536.0){\usebox{\plotpoint}}
\put(437.0,536.0){\rule[-0.200pt]{4.818pt}{0.400pt}}
\put(437.0,537.0){\rule[-0.200pt]{4.818pt}{0.400pt}}
\put(609.0,712.0){\rule[-0.200pt]{0.400pt}{0.723pt}}
\put(599.0,712.0){\rule[-0.200pt]{4.818pt}{0.400pt}}
\put(599.0,715.0){\rule[-0.200pt]{4.818pt}{0.400pt}}
\put(771.0,743.0){\rule[-0.200pt]{0.400pt}{1.445pt}}
\put(761.0,743.0){\rule[-0.200pt]{4.818pt}{0.400pt}}
\put(761.0,749.0){\rule[-0.200pt]{4.818pt}{0.400pt}}
\put(727,230){\makebox(0,0)[r]{$\beta=1.555$}}
\put(771,230){\makebox(0,0){$\triangle$}}
\put(285,327){\makebox(0,0){$\triangle$}}
\put(447,493){\makebox(0,0){$\triangle$}}
\put(609,595){\makebox(0,0){$\triangle$}}
\put(771,734){\makebox(0,0){$\triangle$}}
\put(285.0,324.0){\rule[-0.200pt]{0.400pt}{1.445pt}}
\put(275.0,324.0){\rule[-0.200pt]{4.818pt}{0.400pt}}
\put(275.0,330.0){\rule[-0.200pt]{4.818pt}{0.400pt}}
\put(447.0,485.0){\rule[-0.200pt]{0.400pt}{3.854pt}}
\put(437.0,485.0){\rule[-0.200pt]{4.818pt}{0.400pt}}
\put(437.0,501.0){\rule[-0.200pt]{4.818pt}{0.400pt}}
\put(609.0,578.0){\rule[-0.200pt]{0.400pt}{8.431pt}}
\put(599.0,578.0){\rule[-0.200pt]{4.818pt}{0.400pt}}
\put(599.0,613.0){\rule[-0.200pt]{4.818pt}{0.400pt}}
\put(771.0,635.0){\rule[-0.200pt]{0.400pt}{47.698pt}}
\put(761.0,635.0){\rule[-0.200pt]{4.818pt}{0.400pt}}
\put(761.0,833.0){\rule[-0.200pt]{4.818pt}{0.400pt}}
\put(727,175){\makebox(0,0)[r]{$\beta=1.560$}}
\put(771,175){\circle{24}}
\put(285,308){\circle{24}}
\put(447,456){\circle{24}}
\put(609,565){\circle{24}}
\put(771,600){\circle{24}}
\put(285.0,305.0){\rule[-0.200pt]{0.400pt}{1.445pt}}
\put(275.0,305.0){\rule[-0.200pt]{4.818pt}{0.400pt}}
\put(275.0,311.0){\rule[-0.200pt]{4.818pt}{0.400pt}}
\put(447.0,448.0){\rule[-0.200pt]{0.400pt}{4.095pt}}
\put(437.0,448.0){\rule[-0.200pt]{4.818pt}{0.400pt}}
\put(437.0,465.0){\rule[-0.200pt]{4.818pt}{0.400pt}}
\put(609.0,543.0){\rule[-0.200pt]{0.400pt}{10.600pt}}
\put(599.0,543.0){\rule[-0.200pt]{4.818pt}{0.400pt}}
\put(599.0,587.0){\rule[-0.200pt]{4.818pt}{0.400pt}}
\put(771.0,564.0){\rule[-0.200pt]{0.400pt}{17.345pt}}
\put(761.0,564.0){\rule[-0.200pt]{4.818pt}{0.400pt}}
\put(761.0,636.0){\rule[-0.200pt]{4.818pt}{0.400pt}}
\sbox{\plotpoint}{\rule[-0.500pt]{1.000pt}{1.000pt}}%
\put(220.00,239.00){\usebox{\plotpoint}}
\put(226.96,258.55){\usebox{\plotpoint}}
\put(234.24,277.99){\usebox{\plotpoint}}
\multiput(235,280)(8.176,19.077){0}{\usebox{\plotpoint}}
\put(242.60,296.97){\usebox{\plotpoint}}
\put(251.92,315.50){\usebox{\plotpoint}}
\multiput(254,320)(9.282,18.564){0}{\usebox{\plotpoint}}
\put(261.07,334.13){\usebox{\plotpoint}}
\put(270.66,352.54){\usebox{\plotpoint}}
\multiput(272,355)(9.939,18.221){0}{\usebox{\plotpoint}}
\put(280.91,370.57){\usebox{\plotpoint}}
\multiput(285,377)(10.679,17.798){0}{\usebox{\plotpoint}}
\put(291.82,388.23){\usebox{\plotpoint}}
\put(302.87,405.79){\usebox{\plotpoint}}
\multiput(303,406)(11.513,17.270){0}{\usebox{\plotpoint}}
\put(314.37,423.06){\usebox{\plotpoint}}
\multiput(315,424)(11.513,17.270){0}{\usebox{\plotpoint}}
\put(326.41,439.96){\usebox{\plotpoint}}
\multiput(328,442)(11.513,17.270){0}{\usebox{\plotpoint}}
\put(338.41,456.88){\usebox{\plotpoint}}
\multiput(340,459)(12.453,16.604){0}{\usebox{\plotpoint}}
\put(350.50,473.74){\usebox{\plotpoint}}
\multiput(352,476)(12.453,16.604){0}{\usebox{\plotpoint}}
\put(363.30,490.05){\usebox{\plotpoint}}
\multiput(365,492)(12.453,16.604){0}{\usebox{\plotpoint}}
\put(376.32,506.20){\usebox{\plotpoint}}
\multiput(377,507)(12.453,16.604){0}{\usebox{\plotpoint}}
\put(388.82,522.76){\usebox{\plotpoint}}
\multiput(389,523)(13.508,15.759){0}{\usebox{\plotpoint}}
\multiput(395,530)(13.668,15.620){0}{\usebox{\plotpoint}}
\put(402.40,538.46){\usebox{\plotpoint}}
\multiput(408,545)(13.508,15.759){0}{\usebox{\plotpoint}}
\put(415.76,554.34){\usebox{\plotpoint}}
\multiput(420,560)(13.508,15.759){0}{\usebox{\plotpoint}}
\put(428.90,570.39){\usebox{\plotpoint}}
\multiput(432,574)(14.676,14.676){0}{\usebox{\plotpoint}}
\put(442.97,585.63){\usebox{\plotpoint}}
\multiput(445,588)(13.508,15.759){0}{\usebox{\plotpoint}}
\put(456.48,601.39){\usebox{\plotpoint}}
\multiput(457,602)(13.508,15.759){0}{\usebox{\plotpoint}}
\multiput(463,609)(13.508,15.759){0}{\usebox{\plotpoint}}
\put(469.98,617.15){\usebox{\plotpoint}}
\multiput(475,623)(14.676,14.676){0}{\usebox{\plotpoint}}
\put(484.05,632.39){\usebox{\plotpoint}}
\multiput(488,637)(13.508,15.759){0}{\usebox{\plotpoint}}
\put(497.86,647.86){\usebox{\plotpoint}}
\multiput(500,650)(13.508,15.759){0}{\usebox{\plotpoint}}
\put(511.54,663.47){\usebox{\plotpoint}}
\multiput(512,664)(15.759,13.508){0}{\usebox{\plotpoint}}
\multiput(519,670)(13.508,15.759){0}{\usebox{\plotpoint}}
\put(526.05,678.22){\usebox{\plotpoint}}
\multiput(531,684)(14.676,14.676){0}{\usebox{\plotpoint}}
\put(540.04,693.54){\usebox{\plotpoint}}
\multiput(543,697)(14.676,14.676){0}{\usebox{\plotpoint}}
\put(554.45,708.45){\usebox{\plotpoint}}
\multiput(556,710)(14.676,14.676){0}{\usebox{\plotpoint}}
\multiput(562,716)(14.676,14.676){0}{\usebox{\plotpoint}}
\put(569.04,723.21){\usebox{\plotpoint}}
\multiput(574,729)(14.676,14.676){0}{\usebox{\plotpoint}}
\put(583.29,738.29){\usebox{\plotpoint}}
\multiput(586,741)(15.759,13.508){0}{\usebox{\plotpoint}}
\put(598.45,752.45){\usebox{\plotpoint}}
\multiput(599,753)(14.676,14.676){0}{\usebox{\plotpoint}}
\multiput(605,759)(14.676,14.676){0}{\usebox{\plotpoint}}
\put(613.12,767.12){\usebox{\plotpoint}}
\multiput(617,771)(14.676,14.676){0}{\usebox{\plotpoint}}
\put(627.80,781.80){\usebox{\plotpoint}}
\multiput(629,783)(15.759,13.508){0}{\usebox{\plotpoint}}
\multiput(636,789)(15.945,13.287){0}{\usebox{\plotpoint}}
\put(643.56,795.30){\usebox{\plotpoint}}
\multiput(648,799)(14.676,14.676){0}{\usebox{\plotpoint}}
\put(658.98,809.15){\usebox{\plotpoint}}
\multiput(660,810)(15.945,13.287){0}{\usebox{\plotpoint}}
\multiput(666,815)(16.889,12.064){0}{\usebox{\plotpoint}}
\put(675.51,821.67){\usebox{\plotpoint}}
\multiput(679,824)(15.945,13.287){0}{\usebox{\plotpoint}}
\multiput(685,829)(17.270,11.513){0}{\usebox{\plotpoint}}
\put(692.28,833.86){\usebox{\plotpoint}}
\multiput(697,837)(17.270,11.513){0}{\usebox{\plotpoint}}
\multiput(703,841)(19.077,8.176){0}{\usebox{\plotpoint}}
\put(710.22,844.14){\usebox{\plotpoint}}
\multiput(716,848)(18.564,9.282){0}{\usebox{\plotpoint}}
\multiput(722,851)(19.690,6.563){0}{\usebox{\plotpoint}}
\put(728.69,853.34){\usebox{\plotpoint}}
\multiput(734,856)(19.690,6.563){0}{\usebox{\plotpoint}}
\multiput(740,858)(20.547,2.935){0}{\usebox{\plotpoint}}
\put(748.35,859.45){\usebox{\plotpoint}}
\multiput(753,861)(20.473,3.412){0}{\usebox{\plotpoint}}
\multiput(759,862)(20.756,0.000){0}{\usebox{\plotpoint}}
\put(768.77,862.00){\usebox{\plotpoint}}
\put(771,862){\usebox{\plotpoint}}
\put(220.00,217.00){\usebox{\plotpoint}}
\put(227.54,236.34){\usebox{\plotpoint}}
\multiput(229,240)(7.708,19.271){0}{\usebox{\plotpoint}}
\put(235.27,255.60){\usebox{\plotpoint}}
\put(244.36,274.25){\usebox{\plotpoint}}
\put(253.85,292.70){\usebox{\plotpoint}}
\multiput(254,293)(9.939,18.221){0}{\usebox{\plotpoint}}
\put(263.78,310.93){\usebox{\plotpoint}}
\multiput(266,315)(10.679,17.798){0}{\usebox{\plotpoint}}
\put(274.29,328.82){\usebox{\plotpoint}}
\multiput(278,335)(12.743,16.383){0}{\usebox{\plotpoint}}
\put(286.11,345.84){\usebox{\plotpoint}}
\multiput(291,354)(11.513,17.270){0}{\usebox{\plotpoint}}
\put(297.26,363.34){\usebox{\plotpoint}}
\multiput(303,371)(11.513,17.270){0}{\usebox{\plotpoint}}
\put(309.22,380.29){\usebox{\plotpoint}}
\multiput(315,388)(12.453,16.604){0}{\usebox{\plotpoint}}
\put(321.74,396.84){\usebox{\plotpoint}}
\multiput(328,404)(13.508,15.759){0}{\usebox{\plotpoint}}
\put(335.22,412.62){\usebox{\plotpoint}}
\multiput(340,419)(13.508,15.759){0}{\usebox{\plotpoint}}
\put(348.32,428.70){\usebox{\plotpoint}}
\multiput(352,433)(12.453,16.604){0}{\usebox{\plotpoint}}
\put(361.60,444.60){\usebox{\plotpoint}}
\multiput(365,448)(13.508,15.759){0}{\usebox{\plotpoint}}
\put(375.76,459.76){\usebox{\plotpoint}}
\multiput(377,461)(13.508,15.759){0}{\usebox{\plotpoint}}
\multiput(383,468)(13.508,15.759){0}{\usebox{\plotpoint}}
\put(389.37,475.43){\usebox{\plotpoint}}
\multiput(395,482)(15.759,13.508){0}{\usebox{\plotpoint}}
\put(403.88,490.19){\usebox{\plotpoint}}
\multiput(408,495)(14.676,14.676){0}{\usebox{\plotpoint}}
\put(418.20,505.20){\usebox{\plotpoint}}
\multiput(420,507)(13.508,15.759){0}{\usebox{\plotpoint}}
\multiput(426,514)(14.676,14.676){0}{\usebox{\plotpoint}}
\put(432.38,520.32){\usebox{\plotpoint}}
\multiput(439,526)(14.676,14.676){0}{\usebox{\plotpoint}}
\put(447.51,534.51){\usebox{\plotpoint}}
\multiput(451,538)(14.676,14.676){0}{\usebox{\plotpoint}}
\put(462.19,549.19){\usebox{\plotpoint}}
\multiput(463,550)(14.676,14.676){0}{\usebox{\plotpoint}}
\multiput(469,556)(14.676,14.676){0}{\usebox{\plotpoint}}
\put(477.00,563.71){\usebox{\plotpoint}}
\multiput(482,568)(14.676,14.676){0}{\usebox{\plotpoint}}
\put(492.02,578.02){\usebox{\plotpoint}}
\multiput(494,580)(14.676,14.676){0}{\usebox{\plotpoint}}
\multiput(500,586)(14.676,14.676){0}{\usebox{\plotpoint}}
\put(506.76,592.63){\usebox{\plotpoint}}
\multiput(512,597)(15.759,13.508){0}{\usebox{\plotpoint}}
\put(522.33,606.33){\usebox{\plotpoint}}
\multiput(525,609)(15.945,13.287){0}{\usebox{\plotpoint}}
\multiput(531,614)(14.676,14.676){0}{\usebox{\plotpoint}}
\put(537.53,620.44){\usebox{\plotpoint}}
\multiput(543,625)(14.676,14.676){0}{\usebox{\plotpoint}}
\put(553.19,633.99){\usebox{\plotpoint}}
\multiput(556,636)(15.945,13.287){0}{\usebox{\plotpoint}}
\multiput(562,641)(14.676,14.676){0}{\usebox{\plotpoint}}
\put(568.77,647.64){\usebox{\plotpoint}}
\multiput(574,652)(15.945,13.287){0}{\usebox{\plotpoint}}
\put(584.71,660.93){\usebox{\plotpoint}}
\multiput(586,662)(15.759,13.508){0}{\usebox{\plotpoint}}
\multiput(593,668)(15.945,13.287){0}{\usebox{\plotpoint}}
\put(600.58,674.31){\usebox{\plotpoint}}
\multiput(605,678)(17.270,11.513){0}{\usebox{\plotpoint}}
\put(616.98,686.98){\usebox{\plotpoint}}
\multiput(617,687)(15.945,13.287){0}{\usebox{\plotpoint}}
\multiput(623,692)(15.945,13.287){0}{\usebox{\plotpoint}}
\put(633.44,699.54){\usebox{\plotpoint}}
\multiput(636,701)(15.945,13.287){0}{\usebox{\plotpoint}}
\multiput(642,706)(17.270,11.513){0}{\usebox{\plotpoint}}
\put(650.32,711.54){\usebox{\plotpoint}}
\multiput(654,714)(17.270,11.513){0}{\usebox{\plotpoint}}
\multiput(660,718)(17.270,11.513){0}{\usebox{\plotpoint}}
\put(667.65,722.95){\usebox{\plotpoint}}
\multiput(673,726)(18.564,9.282){0}{\usebox{\plotpoint}}
\multiput(679,729)(17.270,11.513){0}{\usebox{\plotpoint}}
\put(685.61,733.30){\usebox{\plotpoint}}
\multiput(691,736)(18.564,9.282){0}{\usebox{\plotpoint}}
\multiput(697,739)(18.564,9.282){0}{\usebox{\plotpoint}}
\put(704.20,742.52){\usebox{\plotpoint}}
\multiput(710,745)(19.690,6.563){0}{\usebox{\plotpoint}}
\multiput(716,747)(18.564,9.282){0}{\usebox{\plotpoint}}
\put(723.34,750.45){\usebox{\plotpoint}}
\multiput(728,752)(20.473,3.412){0}{\usebox{\plotpoint}}
\multiput(734,753)(19.690,6.563){0}{\usebox{\plotpoint}}
\put(743.41,755.49){\usebox{\plotpoint}}
\multiput(747,756)(20.473,3.412){0}{\usebox{\plotpoint}}
\multiput(753,757)(20.473,3.412){0}{\usebox{\plotpoint}}
\put(763.96,758.00){\usebox{\plotpoint}}
\multiput(765,758)(20.756,0.000){0}{\usebox{\plotpoint}}
\put(771,758){\usebox{\plotpoint}}
\put(220.00,211.00){\usebox{\plotpoint}}
\put(227.88,230.20){\usebox{\plotpoint}}
\multiput(229,233)(8.698,18.845){0}{\usebox{\plotpoint}}
\put(236.53,249.06){\usebox{\plotpoint}}
\put(246.42,267.30){\usebox{\plotpoint}}
\multiput(248,270)(9.939,18.221){0}{\usebox{\plotpoint}}
\put(256.62,285.37){\usebox{\plotpoint}}
\multiput(260,291)(10.679,17.798){0}{\usebox{\plotpoint}}
\put(267.40,303.10){\usebox{\plotpoint}}
\multiput(272,310)(11.513,17.270){0}{\usebox{\plotpoint}}
\put(279.09,320.24){\usebox{\plotpoint}}
\multiput(285,327)(11.513,17.270){0}{\usebox{\plotpoint}}
\put(291.62,336.73){\usebox{\plotpoint}}
\multiput(297,343)(12.453,16.604){0}{\usebox{\plotpoint}}
\put(304.62,352.89){\usebox{\plotpoint}}
\multiput(309,358)(12.453,16.604){0}{\usebox{\plotpoint}}
\put(317.62,369.06){\usebox{\plotpoint}}
\multiput(321,373)(15.759,13.508){0}{\usebox{\plotpoint}}
\put(332.13,383.82){\usebox{\plotpoint}}
\multiput(334,386)(14.676,14.676){0}{\usebox{\plotpoint}}
\multiput(340,392)(13.508,15.759){0}{\usebox{\plotpoint}}
\put(346.12,399.12){\usebox{\plotpoint}}
\multiput(352,405)(14.676,14.676){0}{\usebox{\plotpoint}}
\put(361.01,413.58){\usebox{\plotpoint}}
\multiput(365,417)(14.676,14.676){0}{\usebox{\plotpoint}}
\put(375.96,427.96){\usebox{\plotpoint}}
\multiput(377,429)(14.676,14.676){0}{\usebox{\plotpoint}}
\multiput(383,435)(15.945,13.287){0}{\usebox{\plotpoint}}
\put(391.11,442.11){\usebox{\plotpoint}}
\multiput(395,446)(16.889,12.064){0}{\usebox{\plotpoint}}
\put(406.71,455.71){\usebox{\plotpoint}}
\multiput(408,457)(15.945,13.287){0}{\usebox{\plotpoint}}
\multiput(414,462)(15.945,13.287){0}{\usebox{\plotpoint}}
\put(422.34,469.34){\usebox{\plotpoint}}
\multiput(426,473)(15.945,13.287){0}{\usebox{\plotpoint}}
\put(438.32,482.51){\usebox{\plotpoint}}
\multiput(439,483)(15.945,13.287){0}{\usebox{\plotpoint}}
\multiput(445,488)(15.945,13.287){0}{\usebox{\plotpoint}}
\put(454.30,495.75){\usebox{\plotpoint}}
\multiput(457,498)(15.945,13.287){0}{\usebox{\plotpoint}}
\multiput(463,503)(15.945,13.287){0}{\usebox{\plotpoint}}
\put(470.25,509.04){\usebox{\plotpoint}}
\multiput(475,513)(16.889,12.064){0}{\usebox{\plotpoint}}
\put(486.96,521.31){\usebox{\plotpoint}}
\multiput(488,522)(15.945,13.287){0}{\usebox{\plotpoint}}
\multiput(494,527)(15.945,13.287){0}{\usebox{\plotpoint}}
\put(503.24,534.16){\usebox{\plotpoint}}
\multiput(506,536)(15.945,13.287){0}{\usebox{\plotpoint}}
\multiput(512,541)(18.021,10.298){0}{\usebox{\plotpoint}}
\put(520.20,546.00){\usebox{\plotpoint}}
\multiput(525,550)(17.270,11.513){0}{\usebox{\plotpoint}}
\put(536.60,558.67){\usebox{\plotpoint}}
\multiput(537,559)(17.270,11.513){0}{\usebox{\plotpoint}}
\multiput(543,563)(17.270,11.513){0}{\usebox{\plotpoint}}
\put(553.73,570.38){\usebox{\plotpoint}}
\multiput(556,572)(17.270,11.513){0}{\usebox{\plotpoint}}
\multiput(562,576)(17.270,11.513){0}{\usebox{\plotpoint}}
\put(570.95,581.97){\usebox{\plotpoint}}
\multiput(574,584)(17.270,11.513){0}{\usebox{\plotpoint}}
\multiput(580,588)(17.270,11.513){0}{\usebox{\plotpoint}}
\put(588.32,593.33){\usebox{\plotpoint}}
\multiput(593,596)(17.270,11.513){0}{\usebox{\plotpoint}}
\multiput(599,600)(18.564,9.282){0}{\usebox{\plotpoint}}
\put(606.20,603.80){\usebox{\plotpoint}}
\multiput(611,607)(17.270,11.513){0}{\usebox{\plotpoint}}
\multiput(617,611)(18.564,9.282){0}{\usebox{\plotpoint}}
\put(623.89,614.59){\usebox{\plotpoint}}
\multiput(629,618)(19.077,8.176){0}{\usebox{\plotpoint}}
\multiput(636,621)(18.564,9.282){0}{\usebox{\plotpoint}}
\put(642.26,624.13){\usebox{\plotpoint}}
\multiput(648,627)(18.564,9.282){0}{\usebox{\plotpoint}}
\multiput(654,630)(18.564,9.282){0}{\usebox{\plotpoint}}
\put(660.82,633.41){\usebox{\plotpoint}}
\multiput(666,636)(19.077,8.176){0}{\usebox{\plotpoint}}
\multiput(673,639)(19.690,6.563){0}{\usebox{\plotpoint}}
\put(679.92,641.46){\usebox{\plotpoint}}
\multiput(685,644)(19.690,6.563){0}{\usebox{\plotpoint}}
\multiput(691,646)(19.690,6.563){0}{\usebox{\plotpoint}}
\put(699.30,648.77){\usebox{\plotpoint}}
\multiput(703,650)(19.957,5.702){0}{\usebox{\plotpoint}}
\multiput(710,652)(19.690,6.563){0}{\usebox{\plotpoint}}
\put(719.21,654.53){\usebox{\plotpoint}}
\multiput(722,655)(20.473,3.412){0}{\usebox{\plotpoint}}
\multiput(728,656)(19.690,6.563){0}{\usebox{\plotpoint}}
\put(739.44,658.91){\usebox{\plotpoint}}
\multiput(740,659)(20.756,0.000){0}{\usebox{\plotpoint}}
\multiput(747,659)(20.473,3.412){0}{\usebox{\plotpoint}}
\multiput(753,660)(20.756,0.000){0}{\usebox{\plotpoint}}
\put(760.09,660.18){\usebox{\plotpoint}}
\multiput(765,661)(20.756,0.000){0}{\usebox{\plotpoint}}
\put(771,661){\usebox{\plotpoint}}
\put(220.00,196.00){\usebox{\plotpoint}}
\put(228.18,215.08){\usebox{\plotpoint}}
\multiput(229,217)(8.698,18.845){0}{\usebox{\plotpoint}}
\put(236.94,233.89){\usebox{\plotpoint}}
\put(247.27,251.86){\usebox{\plotpoint}}
\multiput(248,253)(10.679,17.798){0}{\usebox{\plotpoint}}
\put(257.98,269.64){\usebox{\plotpoint}}
\multiput(260,273)(11.513,17.270){0}{\usebox{\plotpoint}}
\put(269.34,287.01){\usebox{\plotpoint}}
\multiput(272,291)(11.513,17.270){0}{\usebox{\plotpoint}}
\put(281.38,303.87){\usebox{\plotpoint}}
\multiput(285,308)(13.508,15.759){0}{\usebox{\plotpoint}}
\put(294.63,319.84){\usebox{\plotpoint}}
\multiput(297,323)(13.508,15.759){0}{\usebox{\plotpoint}}
\put(307.93,335.76){\usebox{\plotpoint}}
\multiput(309,337)(14.676,14.676){0}{\usebox{\plotpoint}}
\multiput(315,343)(13.508,15.759){0}{\usebox{\plotpoint}}
\put(322.07,350.92){\usebox{\plotpoint}}
\multiput(328,356)(14.676,14.676){0}{\usebox{\plotpoint}}
\put(337.16,365.16){\usebox{\plotpoint}}
\multiput(340,368)(14.676,14.676){0}{\usebox{\plotpoint}}
\put(351.83,379.83){\usebox{\plotpoint}}
\multiput(352,380)(14.676,14.676){0}{\usebox{\plotpoint}}
\multiput(358,386)(16.889,12.064){0}{\usebox{\plotpoint}}
\put(367.64,393.20){\usebox{\plotpoint}}
\multiput(371,396)(14.676,14.676){0}{\usebox{\plotpoint}}
\multiput(377,402)(15.945,13.287){0}{\usebox{\plotpoint}}
\put(383.06,407.05){\usebox{\plotpoint}}
\multiput(389,412)(15.945,13.287){0}{\usebox{\plotpoint}}
\put(399.24,420.03){\usebox{\plotpoint}}
\multiput(402,422)(15.945,13.287){0}{\usebox{\plotpoint}}
\multiput(408,427)(15.945,13.287){0}{\usebox{\plotpoint}}
\put(415.34,433.12){\usebox{\plotpoint}}
\multiput(420,437)(17.270,11.513){0}{\usebox{\plotpoint}}
\put(431.75,445.79){\usebox{\plotpoint}}
\multiput(432,446)(16.889,12.064){0}{\usebox{\plotpoint}}
\multiput(439,451)(17.270,11.513){0}{\usebox{\plotpoint}}
\put(448.54,457.95){\usebox{\plotpoint}}
\multiput(451,460)(17.270,11.513){0}{\usebox{\plotpoint}}
\multiput(457,464)(15.945,13.287){0}{\usebox{\plotpoint}}
\put(465.11,470.41){\usebox{\plotpoint}}
\multiput(469,473)(17.270,11.513){0}{\usebox{\plotpoint}}
\multiput(475,477)(16.889,12.064){0}{\usebox{\plotpoint}}
\put(482.22,482.15){\usebox{\plotpoint}}
\multiput(488,486)(17.270,11.513){0}{\usebox{\plotpoint}}
\put(499.49,493.66){\usebox{\plotpoint}}
\multiput(500,494)(17.270,11.513){0}{\usebox{\plotpoint}}
\multiput(506,498)(17.270,11.513){0}{\usebox{\plotpoint}}
\put(516.97,504.84){\usebox{\plotpoint}}
\multiput(519,506)(17.270,11.513){0}{\usebox{\plotpoint}}
\multiput(525,510)(17.270,11.513){0}{\usebox{\plotpoint}}
\put(534.32,516.22){\usebox{\plotpoint}}
\multiput(537,518)(17.270,11.513){0}{\usebox{\plotpoint}}
\multiput(543,522)(18.564,9.282){0}{\usebox{\plotpoint}}
\put(552.14,526.80){\usebox{\plotpoint}}
\multiput(556,529)(17.270,11.513){0}{\usebox{\plotpoint}}
\multiput(562,533)(18.564,9.282){0}{\usebox{\plotpoint}}
\put(569.99,537.33){\usebox{\plotpoint}}
\multiput(574,540)(18.564,9.282){0}{\usebox{\plotpoint}}
\multiput(580,543)(18.564,9.282){0}{\usebox{\plotpoint}}
\put(588.19,547.25){\usebox{\plotpoint}}
\multiput(593,550)(18.564,9.282){0}{\usebox{\plotpoint}}
\multiput(599,553)(18.564,9.282){0}{\usebox{\plotpoint}}
\put(606.61,556.80){\usebox{\plotpoint}}
\multiput(611,559)(18.564,9.282){0}{\usebox{\plotpoint}}
\multiput(617,562)(18.564,9.282){0}{\usebox{\plotpoint}}
\put(625.17,566.09){\usebox{\plotpoint}}
\multiput(629,568)(19.077,8.176){0}{\usebox{\plotpoint}}
\multiput(636,571)(18.564,9.282){0}{\usebox{\plotpoint}}
\put(644.04,574.68){\usebox{\plotpoint}}
\multiput(648,576)(18.564,9.282){0}{\usebox{\plotpoint}}
\multiput(654,579)(19.690,6.563){0}{\usebox{\plotpoint}}
\put(663.37,582.12){\usebox{\plotpoint}}
\multiput(666,583)(19.077,8.176){0}{\usebox{\plotpoint}}
\multiput(673,586)(19.690,6.563){0}{\usebox{\plotpoint}}
\put(682.83,589.28){\usebox{\plotpoint}}
\multiput(685,590)(19.690,6.563){0}{\usebox{\plotpoint}}
\multiput(691,592)(20.473,3.412){0}{\usebox{\plotpoint}}
\put(702.75,594.92){\usebox{\plotpoint}}
\multiput(703,595)(20.547,2.935){0}{\usebox{\plotpoint}}
\multiput(710,596)(19.690,6.563){0}{\usebox{\plotpoint}}
\multiput(716,598)(20.473,3.412){0}{\usebox{\plotpoint}}
\put(723.00,599.17){\usebox{\plotpoint}}
\multiput(728,600)(20.473,3.412){0}{\usebox{\plotpoint}}
\multiput(734,601)(20.473,3.412){0}{\usebox{\plotpoint}}
\put(743.53,602.00){\usebox{\plotpoint}}
\multiput(747,602)(20.473,3.412){0}{\usebox{\plotpoint}}
\multiput(753,603)(20.756,0.000){0}{\usebox{\plotpoint}}
\put(764.20,603.00){\usebox{\plotpoint}}
\multiput(765,603)(20.473,3.412){0}{\usebox{\plotpoint}}
\put(771,604){\usebox{\plotpoint}}
\end{picture}

%% file: phasefig4b.tex
\setlength{\unitlength}{0.240900pt}
\ifx\plotpoint\undefined\newsavebox{\plotpoint}\fi
\sbox{\plotpoint}{\rule[-0.200pt]{0.400pt}{0.400pt}}%
\begin{picture}(900,900)(0,0)
\font\gnuplot=cmr10 at 10pt
\gnuplot
\sbox{\plotpoint}{\rule[-0.200pt]{0.400pt}{0.400pt}}%
\put(220.0,155.0){\rule[-0.200pt]{4.818pt}{0.400pt}}
\put(198,155){\makebox(0,0)[r]{0.8}}
\put(816.0,155.0){\rule[-0.200pt]{4.818pt}{0.400pt}}
\put(220.0,240.0){\rule[-0.200pt]{4.818pt}{0.400pt}}
\put(198,240){\makebox(0,0)[r]{1}}
\put(816.0,240.0){\rule[-0.200pt]{4.818pt}{0.400pt}}
\put(220.0,325.0){\rule[-0.200pt]{4.818pt}{0.400pt}}
\put(198,325){\makebox(0,0)[r]{1.2}}
\put(816.0,325.0){\rule[-0.200pt]{4.818pt}{0.400pt}}
\put(220.0,410.0){\rule[-0.200pt]{4.818pt}{0.400pt}}
\put(198,410){\makebox(0,0)[r]{1.4}}
\put(816.0,410.0){\rule[-0.200pt]{4.818pt}{0.400pt}}
\put(220.0,495.0){\rule[-0.200pt]{4.818pt}{0.400pt}}
\put(198,495){\makebox(0,0)[r]{1.6}}
\put(816.0,495.0){\rule[-0.200pt]{4.818pt}{0.400pt}}
\put(220.0,580.0){\rule[-0.200pt]{4.818pt}{0.400pt}}
\put(198,580){\makebox(0,0)[r]{1.8}}
\put(816.0,580.0){\rule[-0.200pt]{4.818pt}{0.400pt}}
\put(220.0,665.0){\rule[-0.200pt]{4.818pt}{0.400pt}}
\put(198,665){\makebox(0,0)[r]{2}}
\put(816.0,665.0){\rule[-0.200pt]{4.818pt}{0.400pt}}
\put(220.0,750.0){\rule[-0.200pt]{4.818pt}{0.400pt}}
\put(198,750){\makebox(0,0)[r]{2.2}}
\put(816.0,750.0){\rule[-0.200pt]{4.818pt}{0.400pt}}
\put(220.0,835.0){\rule[-0.200pt]{4.818pt}{0.400pt}}
\put(198,835){\makebox(0,0)[r]{2.4}}
\put(816.0,835.0){\rule[-0.200pt]{4.818pt}{0.400pt}}
\put(285.0,113.0){\rule[-0.200pt]{0.400pt}{4.818pt}}
\put(285,68){\makebox(0,0){1}}
\put(285.0,857.0){\rule[-0.200pt]{0.400pt}{4.818pt}}
\put(447.0,113.0){\rule[-0.200pt]{0.400pt}{4.818pt}}
\put(447,68){\makebox(0,0){2}}
\put(447.0,857.0){\rule[-0.200pt]{0.400pt}{4.818pt}}
\put(609.0,113.0){\rule[-0.200pt]{0.400pt}{4.818pt}}
\put(609,68){\makebox(0,0){3}}
\put(609.0,857.0){\rule[-0.200pt]{0.400pt}{4.818pt}}
\put(771.0,113.0){\rule[-0.200pt]{0.400pt}{4.818pt}}
\put(771,68){\makebox(0,0){4}}
\put(771.0,857.0){\rule[-0.200pt]{0.400pt}{4.818pt}}
\put(220.0,113.0){\rule[-0.200pt]{148.394pt}{0.400pt}}
\put(836.0,113.0){\rule[-0.200pt]{0.400pt}{184.048pt}}
\put(220.0,877.0){\rule[-0.200pt]{148.394pt}{0.400pt}}
\put(60,950){\makebox(0,0){(b)}}
\put(45,495){\makebox(0,0){$V(R)$}}
\put(528,10){\makebox(0,0){$R$}}
\put(220.0,113.0){\rule[-0.200pt]{0.400pt}{184.048pt}}
\put(727,395){\makebox(0,0)[r]{$\beta=1.020$}}
\put(771,395){\raisebox{-.8pt}{\makebox(0,0){$\Diamond$}}}
\put(285,289){\raisebox{-.8pt}{\makebox(0,0){$\Diamond$}}}
\put(447,511){\raisebox{-.8pt}{\makebox(0,0){$\Diamond$}}}
\put(609,681){\raisebox{-.8pt}{\makebox(0,0){$\Diamond$}}}
\put(771,789){\raisebox{-.8pt}{\makebox(0,0){$\Diamond$}}}
\put(285.0,289.0){\usebox{\plotpoint}}
\put(275.0,289.0){\rule[-0.200pt]{4.818pt}{0.400pt}}
\put(275.0,290.0){\rule[-0.200pt]{4.818pt}{0.400pt}}
\put(447.0,508.0){\rule[-0.200pt]{0.400pt}{1.686pt}}
\put(437.0,508.0){\rule[-0.200pt]{4.818pt}{0.400pt}}
\put(437.0,515.0){\rule[-0.200pt]{4.818pt}{0.400pt}}
\put(609.0,665.0){\rule[-0.200pt]{0.400pt}{7.950pt}}
\put(599.0,665.0){\rule[-0.200pt]{4.818pt}{0.400pt}}
\put(599.0,698.0){\rule[-0.200pt]{4.818pt}{0.400pt}}
\put(771.0,724.0){\rule[-0.200pt]{0.400pt}{31.076pt}}
\put(761.0,724.0){\rule[-0.200pt]{4.818pt}{0.400pt}}
\put(761.0,853.0){\rule[-0.200pt]{4.818pt}{0.400pt}}
\put(727,340){\makebox(0,0)[r]{$\beta=1.025$}}
\put(771,340){\raisebox{-.8pt}{\makebox(0,0){$\Box$}}}
\put(285,281){\raisebox{-.8pt}{\makebox(0,0){$\Box$}}}
\put(447,503){\raisebox{-.8pt}{\makebox(0,0){$\Box$}}}
\put(609,629){\raisebox{-.8pt}{\makebox(0,0){$\Box$}}}
\put(771,715){\raisebox{-.8pt}{\makebox(0,0){$\Box$}}}
\put(285.0,280.0){\rule[-0.200pt]{0.400pt}{0.482pt}}
\put(275.0,280.0){\rule[-0.200pt]{4.818pt}{0.400pt}}
\put(275.0,282.0){\rule[-0.200pt]{4.818pt}{0.400pt}}
\put(447.0,494.0){\rule[-0.200pt]{0.400pt}{4.095pt}}
\put(437.0,494.0){\rule[-0.200pt]{4.818pt}{0.400pt}}
\put(437.0,511.0){\rule[-0.200pt]{4.818pt}{0.400pt}}
\put(609.0,595.0){\rule[-0.200pt]{0.400pt}{16.381pt}}
\put(599.0,595.0){\rule[-0.200pt]{4.818pt}{0.400pt}}
\put(599.0,663.0){\rule[-0.200pt]{4.818pt}{0.400pt}}
\put(771.0,647.0){\rule[-0.200pt]{0.400pt}{33.003pt}}
\put(761.0,647.0){\rule[-0.200pt]{4.818pt}{0.400pt}}
\put(761.0,784.0){\rule[-0.200pt]{4.818pt}{0.400pt}}
\put(727,285){\makebox(0,0)[r]{$\beta=1.030$}}
\put(771,285){\makebox(0,0){$\triangle$}}
\put(285,269){\makebox(0,0){$\triangle$}}
\put(447,457){\makebox(0,0){$\triangle$}}
\put(609,602){\makebox(0,0){$\triangle$}}
\put(771,742){\makebox(0,0){$\triangle$}}
\put(285.0,268.0){\rule[-0.200pt]{0.400pt}{0.482pt}}
\put(275.0,268.0){\rule[-0.200pt]{4.818pt}{0.400pt}}
\put(275.0,270.0){\rule[-0.200pt]{4.818pt}{0.400pt}}
\put(447.0,452.0){\rule[-0.200pt]{0.400pt}{2.650pt}}
\put(437.0,452.0){\rule[-0.200pt]{4.818pt}{0.400pt}}
\put(437.0,463.0){\rule[-0.200pt]{4.818pt}{0.400pt}}
\put(609.0,581.0){\rule[-0.200pt]{0.400pt}{10.118pt}}
\put(599.0,581.0){\rule[-0.200pt]{4.818pt}{0.400pt}}
\put(599.0,623.0){\rule[-0.200pt]{4.818pt}{0.400pt}}
\put(771.0,652.0){\rule[-0.200pt]{0.400pt}{43.362pt}}
\put(761.0,652.0){\rule[-0.200pt]{4.818pt}{0.400pt}}
\put(761.0,832.0){\rule[-0.200pt]{4.818pt}{0.400pt}}
\put(727,230){\makebox(0,0)[r]{$\beta=1.035$}}
\put(771,230){\makebox(0,0){$\star$}}
\put(285,258){\makebox(0,0){$\star$}}
\put(447,433){\makebox(0,0){$\star$}}
\put(609,521){\makebox(0,0){$\star$}}
\put(771,615){\makebox(0,0){$\star$}}
\put(285.0,257.0){\rule[-0.200pt]{0.400pt}{0.482pt}}
\put(275.0,257.0){\rule[-0.200pt]{4.818pt}{0.400pt}}
\put(275.0,259.0){\rule[-0.200pt]{4.818pt}{0.400pt}}
\put(447.0,428.0){\rule[-0.200pt]{0.400pt}{2.409pt}}
\put(437.0,428.0){\rule[-0.200pt]{4.818pt}{0.400pt}}
\put(437.0,438.0){\rule[-0.200pt]{4.818pt}{0.400pt}}
\put(609.0,508.0){\rule[-0.200pt]{0.400pt}{6.504pt}}
\put(599.0,508.0){\rule[-0.200pt]{4.818pt}{0.400pt}}
\put(599.0,535.0){\rule[-0.200pt]{4.818pt}{0.400pt}}
\put(771.0,574.0){\rule[-0.200pt]{0.400pt}{19.995pt}}
\put(761.0,574.0){\rule[-0.200pt]{4.818pt}{0.400pt}}
\put(761.0,657.0){\rule[-0.200pt]{4.818pt}{0.400pt}}
\put(727,175){\makebox(0,0)[r]{$\beta=1.040$}}
\put(771,175){\circle{24}}
\put(285,250){\circle{24}}
\put(447,411){\circle{24}}
\put(609,526){\circle{24}}
\put(771,549){\circle{24}}
\put(285.0,249.0){\rule[-0.200pt]{0.400pt}{0.482pt}}
\put(275.0,249.0){\rule[-0.200pt]{4.818pt}{0.400pt}}
\put(275.0,251.0){\rule[-0.200pt]{4.818pt}{0.400pt}}
\put(447.0,406.0){\rule[-0.200pt]{0.400pt}{2.650pt}}
\put(437.0,406.0){\rule[-0.200pt]{4.818pt}{0.400pt}}
\put(437.0,417.0){\rule[-0.200pt]{4.818pt}{0.400pt}}
\put(609.0,511.0){\rule[-0.200pt]{0.400pt}{7.227pt}}
\put(599.0,511.0){\rule[-0.200pt]{4.818pt}{0.400pt}}
\put(599.0,541.0){\rule[-0.200pt]{4.818pt}{0.400pt}}
\put(771.0,526.0){\rule[-0.200pt]{0.400pt}{11.081pt}}
\put(761.0,526.0){\rule[-0.200pt]{4.818pt}{0.400pt}}
\put(761.0,572.0){\rule[-0.200pt]{4.818pt}{0.400pt}}
\sbox{\plotpoint}{\rule[-0.500pt]{1.000pt}{1.000pt}}%
\put(220.00,148.00){\usebox{\plotpoint}}
\put(226.75,167.63){\usebox{\plotpoint}}
\put(234.20,187.00){\usebox{\plotpoint}}
\multiput(235,189)(7.708,19.271){0}{\usebox{\plotpoint}}
\put(242.09,206.18){\usebox{\plotpoint}}
\put(251.16,224.85){\usebox{\plotpoint}}
\multiput(254,231)(9.282,18.564){0}{\usebox{\plotpoint}}
\put(260.25,243.51){\usebox{\plotpoint}}
\put(269.53,262.07){\usebox{\plotpoint}}
\multiput(272,267)(9.939,18.221){0}{\usebox{\plotpoint}}
\put(279.46,280.29){\usebox{\plotpoint}}
\put(290.37,297.94){\usebox{\plotpoint}}
\multiput(291,299)(10.679,17.798){0}{\usebox{\plotpoint}}
\put(301.05,315.74){\usebox{\plotpoint}}
\multiput(303,319)(10.679,17.798){0}{\usebox{\plotpoint}}
\put(311.94,333.40){\usebox{\plotpoint}}
\multiput(315,338)(11.513,17.270){0}{\usebox{\plotpoint}}
\put(323.71,350.49){\usebox{\plotpoint}}
\multiput(328,356)(11.513,17.270){0}{\usebox{\plotpoint}}
\put(335.64,367.46){\usebox{\plotpoint}}
\multiput(340,374)(12.453,16.604){0}{\usebox{\plotpoint}}
\put(347.60,384.41){\usebox{\plotpoint}}
\multiput(352,391)(12.453,16.604){0}{\usebox{\plotpoint}}
\put(359.74,401.23){\usebox{\plotpoint}}
\multiput(365,408)(12.453,16.604){0}{\usebox{\plotpoint}}
\put(372.31,417.75){\usebox{\plotpoint}}
\multiput(377,424)(12.453,16.604){0}{\usebox{\plotpoint}}
\put(384.76,434.35){\usebox{\plotpoint}}
\multiput(389,440)(12.453,16.604){0}{\usebox{\plotpoint}}
\put(397.61,450.61){\usebox{\plotpoint}}
\multiput(402,455)(12.453,16.604){0}{\usebox{\plotpoint}}
\put(410.73,466.64){\usebox{\plotpoint}}
\multiput(414,471)(13.508,15.759){0}{\usebox{\plotpoint}}
\put(423.65,482.87){\usebox{\plotpoint}}
\multiput(426,486)(13.508,15.759){0}{\usebox{\plotpoint}}
\put(437.02,498.74){\usebox{\plotpoint}}
\multiput(439,501)(13.508,15.759){0}{\usebox{\plotpoint}}
\put(450.55,514.48){\usebox{\plotpoint}}
\multiput(451,515)(12.453,16.604){0}{\usebox{\plotpoint}}
\multiput(457,523)(13.508,15.759){0}{\usebox{\plotpoint}}
\put(463.55,530.64){\usebox{\plotpoint}}
\multiput(469,537)(13.508,15.759){0}{\usebox{\plotpoint}}
\put(477.08,546.38){\usebox{\plotpoint}}
\multiput(482,552)(13.508,15.759){0}{\usebox{\plotpoint}}
\put(490.65,562.09){\usebox{\plotpoint}}
\multiput(494,566)(13.508,15.759){0}{\usebox{\plotpoint}}
\put(504.15,577.85){\usebox{\plotpoint}}
\multiput(506,580)(13.508,15.759){0}{\usebox{\plotpoint}}
\put(518.15,593.15){\usebox{\plotpoint}}
\multiput(519,594)(13.508,15.759){0}{\usebox{\plotpoint}}
\multiput(525,601)(13.508,15.759){0}{\usebox{\plotpoint}}
\put(531.73,608.85){\usebox{\plotpoint}}
\multiput(537,615)(14.676,14.676){0}{\usebox{\plotpoint}}
\put(545.71,624.16){\usebox{\plotpoint}}
\multiput(549,628)(14.676,14.676){0}{\usebox{\plotpoint}}
\put(559.78,639.41){\usebox{\plotpoint}}
\multiput(562,642)(14.676,14.676){0}{\usebox{\plotpoint}}
\put(573.76,654.72){\usebox{\plotpoint}}
\multiput(574,655)(13.508,15.759){0}{\usebox{\plotpoint}}
\multiput(580,662)(14.676,14.676){0}{\usebox{\plotpoint}}
\put(587.90,669.90){\usebox{\plotpoint}}
\multiput(593,675)(14.676,14.676){0}{\usebox{\plotpoint}}
\put(602.58,684.58){\usebox{\plotpoint}}
\multiput(605,687)(13.508,15.759){0}{\usebox{\plotpoint}}
\put(616.73,699.73){\usebox{\plotpoint}}
\multiput(617,700)(14.676,14.676){0}{\usebox{\plotpoint}}
\multiput(623,706)(14.676,14.676){0}{\usebox{\plotpoint}}
\put(631.59,714.22){\usebox{\plotpoint}}
\multiput(636,718)(14.676,14.676){0}{\usebox{\plotpoint}}
\put(646.57,728.57){\usebox{\plotpoint}}
\multiput(648,730)(15.945,13.287){0}{\usebox{\plotpoint}}
\multiput(654,735)(14.676,14.676){0}{\usebox{\plotpoint}}
\put(661.87,742.56){\usebox{\plotpoint}}
\multiput(666,746)(15.759,13.508){0}{\usebox{\plotpoint}}
\put(677.73,755.94){\usebox{\plotpoint}}
\multiput(679,757)(17.270,11.513){0}{\usebox{\plotpoint}}
\multiput(685,761)(15.945,13.287){0}{\usebox{\plotpoint}}
\put(694.40,768.26){\usebox{\plotpoint}}
\multiput(697,770)(17.270,11.513){0}{\usebox{\plotpoint}}
\multiput(703,774)(18.021,10.298){0}{\usebox{\plotpoint}}
\put(711.96,779.31){\usebox{\plotpoint}}
\multiput(716,782)(18.564,9.282){0}{\usebox{\plotpoint}}
\multiput(722,785)(18.564,9.282){0}{\usebox{\plotpoint}}
\put(730.35,788.78){\usebox{\plotpoint}}
\multiput(734,790)(18.564,9.282){0}{\usebox{\plotpoint}}
\multiput(740,793)(20.547,2.935){0}{\usebox{\plotpoint}}
\put(749.97,794.99){\usebox{\plotpoint}}
\multiput(753,796)(20.473,3.412){0}{\usebox{\plotpoint}}
\multiput(759,797)(20.756,0.000){0}{\usebox{\plotpoint}}
\put(770.41,797.90){\usebox{\plotpoint}}
\put(771,798){\usebox{\plotpoint}}
\put(220.00,141.00){\usebox{\plotpoint}}
\put(227.06,160.52){\usebox{\plotpoint}}
\put(234.25,179.99){\usebox{\plotpoint}}
\multiput(235,182)(8.176,19.077){0}{\usebox{\plotpoint}}
\put(242.51,199.02){\usebox{\plotpoint}}
\put(251.55,217.70){\usebox{\plotpoint}}
\multiput(254,223)(8.698,18.845){0}{\usebox{\plotpoint}}
\put(260.27,236.53){\usebox{\plotpoint}}
\put(269.80,254.97){\usebox{\plotpoint}}
\multiput(272,259)(9.939,18.221){0}{\usebox{\plotpoint}}
\put(279.95,273.06){\usebox{\plotpoint}}
\put(290.84,290.73){\usebox{\plotpoint}}
\multiput(291,291)(9.939,18.221){0}{\usebox{\plotpoint}}
\put(301.39,308.58){\usebox{\plotpoint}}
\multiput(303,311)(10.679,17.798){0}{\usebox{\plotpoint}}
\put(312.18,326.31){\usebox{\plotpoint}}
\multiput(315,331)(11.513,17.270){0}{\usebox{\plotpoint}}
\put(323.74,343.53){\usebox{\plotpoint}}
\multiput(328,349)(11.513,17.270){0}{\usebox{\plotpoint}}
\put(335.80,360.40){\usebox{\plotpoint}}
\multiput(340,366)(11.513,17.270){0}{\usebox{\plotpoint}}
\put(347.63,377.45){\usebox{\plotpoint}}
\multiput(352,384)(12.453,16.604){0}{\usebox{\plotpoint}}
\put(359.90,394.17){\usebox{\plotpoint}}
\multiput(365,400)(12.453,16.604){0}{\usebox{\plotpoint}}
\put(372.80,410.41){\usebox{\plotpoint}}
\multiput(377,416)(12.453,16.604){0}{\usebox{\plotpoint}}
\put(385.26,427.01){\usebox{\plotpoint}}
\multiput(389,432)(12.453,16.604){0}{\usebox{\plotpoint}}
\put(397.98,443.40){\usebox{\plotpoint}}
\multiput(402,448)(12.453,16.604){0}{\usebox{\plotpoint}}
\put(411.02,459.53){\usebox{\plotpoint}}
\multiput(414,463)(12.453,16.604){0}{\usebox{\plotpoint}}
\put(424.02,475.69){\usebox{\plotpoint}}
\multiput(426,478)(12.453,16.604){0}{\usebox{\plotpoint}}
\put(437.46,491.46){\usebox{\plotpoint}}
\multiput(439,493)(12.453,16.604){0}{\usebox{\plotpoint}}
\put(450.58,507.51){\usebox{\plotpoint}}
\multiput(451,508)(13.508,15.759){0}{\usebox{\plotpoint}}
\multiput(457,515)(12.453,16.604){0}{\usebox{\plotpoint}}
\put(463.58,523.67){\usebox{\plotpoint}}
\multiput(469,530)(13.508,15.759){0}{\usebox{\plotpoint}}
\put(477.27,539.27){\usebox{\plotpoint}}
\multiput(482,544)(13.508,15.759){0}{\usebox{\plotpoint}}
\put(490.90,554.87){\usebox{\plotpoint}}
\multiput(494,559)(13.508,15.759){0}{\usebox{\plotpoint}}
\put(504.15,570.84){\usebox{\plotpoint}}
\multiput(506,573)(13.508,15.759){0}{\usebox{\plotpoint}}
\put(518.15,586.15){\usebox{\plotpoint}}
\multiput(519,587)(13.508,15.759){0}{\usebox{\plotpoint}}
\multiput(525,594)(14.676,14.676){0}{\usebox{\plotpoint}}
\put(532.20,601.40){\usebox{\plotpoint}}
\multiput(537,607)(13.508,15.759){0}{\usebox{\plotpoint}}
\put(545.71,617.16){\usebox{\plotpoint}}
\multiput(549,621)(14.676,14.676){0}{\usebox{\plotpoint}}
\put(560.10,632.10){\usebox{\plotpoint}}
\multiput(562,634)(13.508,15.759){0}{\usebox{\plotpoint}}
\put(573.76,647.72){\usebox{\plotpoint}}
\multiput(574,648)(14.676,14.676){0}{\usebox{\plotpoint}}
\multiput(580,654)(13.508,15.759){0}{\usebox{\plotpoint}}
\put(588.03,662.74){\usebox{\plotpoint}}
\multiput(593,667)(13.508,15.759){0}{\usebox{\plotpoint}}
\put(602.53,677.53){\usebox{\plotpoint}}
\multiput(605,680)(13.508,15.759){0}{\usebox{\plotpoint}}
\put(616.69,692.69){\usebox{\plotpoint}}
\multiput(617,693)(14.676,14.676){0}{\usebox{\plotpoint}}
\multiput(623,699)(14.676,14.676){0}{\usebox{\plotpoint}}
\put(631.54,707.18){\usebox{\plotpoint}}
\multiput(636,711)(14.676,14.676){0}{\usebox{\plotpoint}}
\put(646.52,721.52){\usebox{\plotpoint}}
\multiput(648,723)(15.945,13.287){0}{\usebox{\plotpoint}}
\multiput(654,728)(14.676,14.676){0}{\usebox{\plotpoint}}
\put(661.82,735.52){\usebox{\plotpoint}}
\multiput(666,739)(16.889,12.064){0}{\usebox{\plotpoint}}
\put(678.16,748.30){\usebox{\plotpoint}}
\multiput(679,749)(15.945,13.287){0}{\usebox{\plotpoint}}
\multiput(685,754)(15.945,13.287){0}{\usebox{\plotpoint}}
\put(694.36,761.24){\usebox{\plotpoint}}
\multiput(697,763)(17.270,11.513){0}{\usebox{\plotpoint}}
\multiput(703,767)(18.021,10.298){0}{\usebox{\plotpoint}}
\put(711.92,772.28){\usebox{\plotpoint}}
\multiput(716,775)(18.564,9.282){0}{\usebox{\plotpoint}}
\multiput(722,778)(18.564,9.282){0}{\usebox{\plotpoint}}
\put(730.31,781.77){\usebox{\plotpoint}}
\multiput(734,783)(18.564,9.282){0}{\usebox{\plotpoint}}
\multiput(740,786)(20.547,2.935){0}{\usebox{\plotpoint}}
\put(749.93,787.98){\usebox{\plotpoint}}
\multiput(753,789)(20.473,3.412){0}{\usebox{\plotpoint}}
\multiput(759,790)(20.756,0.000){0}{\usebox{\plotpoint}}
\put(770.36,790.89){\usebox{\plotpoint}}
\put(771,791){\usebox{\plotpoint}}
\put(220.00,141.00){\usebox{\plotpoint}}
\put(227.54,160.34){\usebox{\plotpoint}}
\multiput(229,164)(7.708,19.271){0}{\usebox{\plotpoint}}
\put(235.27,179.60){\usebox{\plotpoint}}
\put(244.36,198.25){\usebox{\plotpoint}}
\put(253.85,216.70){\usebox{\plotpoint}}
\multiput(254,217)(9.939,18.221){0}{\usebox{\plotpoint}}
\put(263.78,234.93){\usebox{\plotpoint}}
\multiput(266,239)(10.679,17.798){0}{\usebox{\plotpoint}}
\put(274.29,252.82){\usebox{\plotpoint}}
\multiput(278,259)(11.902,17.004){0}{\usebox{\plotpoint}}
\put(285.75,270.12){\usebox{\plotpoint}}
\multiput(291,278)(11.513,17.270){0}{\usebox{\plotpoint}}
\put(297.28,287.37){\usebox{\plotpoint}}
\multiput(303,295)(11.513,17.270){0}{\usebox{\plotpoint}}
\put(309.24,304.32){\usebox{\plotpoint}}
\multiput(315,312)(12.453,16.604){0}{\usebox{\plotpoint}}
\put(321.76,320.87){\usebox{\plotpoint}}
\multiput(328,328)(12.453,16.604){0}{\usebox{\plotpoint}}
\put(334.84,336.98){\usebox{\plotpoint}}
\multiput(340,343)(13.508,15.759){0}{\usebox{\plotpoint}}
\put(348.16,352.88){\usebox{\plotpoint}}
\multiput(352,358)(13.508,15.759){0}{\usebox{\plotpoint}}
\put(361.63,368.63){\usebox{\plotpoint}}
\multiput(365,372)(13.508,15.759){0}{\usebox{\plotpoint}}
\put(375.41,384.14){\usebox{\plotpoint}}
\multiput(377,386)(13.508,15.759){0}{\usebox{\plotpoint}}
\multiput(383,393)(14.676,14.676){0}{\usebox{\plotpoint}}
\put(389.39,399.46){\usebox{\plotpoint}}
\multiput(395,406)(15.759,13.508){0}{\usebox{\plotpoint}}
\put(403.90,414.22){\usebox{\plotpoint}}
\multiput(408,419)(14.676,14.676){0}{\usebox{\plotpoint}}
\put(417.89,429.54){\usebox{\plotpoint}}
\multiput(420,432)(14.676,14.676){0}{\usebox{\plotpoint}}
\multiput(426,438)(14.676,14.676){0}{\usebox{\plotpoint}}
\put(432.38,444.38){\usebox{\plotpoint}}
\multiput(439,451)(14.676,14.676){0}{\usebox{\plotpoint}}
\put(447.06,459.06){\usebox{\plotpoint}}
\multiput(451,463)(14.676,14.676){0}{\usebox{\plotpoint}}
\put(461.73,473.73){\usebox{\plotpoint}}
\multiput(463,475)(14.676,14.676){0}{\usebox{\plotpoint}}
\multiput(469,481)(14.676,14.676){0}{\usebox{\plotpoint}}
\put(476.51,488.30){\usebox{\plotpoint}}
\multiput(482,493)(14.676,14.676){0}{\usebox{\plotpoint}}
\put(491.57,502.57){\usebox{\plotpoint}}
\multiput(494,505)(15.945,13.287){0}{\usebox{\plotpoint}}
\multiput(500,510)(14.676,14.676){0}{\usebox{\plotpoint}}
\put(506.72,516.72){\usebox{\plotpoint}}
\multiput(512,522)(15.759,13.508){0}{\usebox{\plotpoint}}
\put(522.13,530.61){\usebox{\plotpoint}}
\multiput(525,533)(14.676,14.676){0}{\usebox{\plotpoint}}
\multiput(531,539)(15.945,13.287){0}{\usebox{\plotpoint}}
\put(537.51,544.51){\usebox{\plotpoint}}
\multiput(543,550)(15.945,13.287){0}{\usebox{\plotpoint}}
\put(552.93,558.37){\usebox{\plotpoint}}
\multiput(556,561)(15.945,13.287){0}{\usebox{\plotpoint}}
\multiput(562,566)(14.676,14.676){0}{\usebox{\plotpoint}}
\put(568.32,572.27){\usebox{\plotpoint}}
\multiput(574,577)(15.945,13.287){0}{\usebox{\plotpoint}}
\put(584.27,585.56){\usebox{\plotpoint}}
\multiput(586,587)(16.889,12.064){0}{\usebox{\plotpoint}}
\multiput(593,592)(15.945,13.287){0}{\usebox{\plotpoint}}
\put(600.60,598.34){\usebox{\plotpoint}}
\multiput(605,602)(15.945,13.287){0}{\usebox{\plotpoint}}
\put(616.55,611.62){\usebox{\plotpoint}}
\multiput(617,612)(15.945,13.287){0}{\usebox{\plotpoint}}
\multiput(623,617)(15.945,13.287){0}{\usebox{\plotpoint}}
\put(632.95,624.26){\usebox{\plotpoint}}
\multiput(636,626)(15.945,13.287){0}{\usebox{\plotpoint}}
\multiput(642,631)(17.270,11.513){0}{\usebox{\plotpoint}}
\put(649.85,636.23){\usebox{\plotpoint}}
\multiput(654,639)(17.270,11.513){0}{\usebox{\plotpoint}}
\multiput(660,643)(17.270,11.513){0}{\usebox{\plotpoint}}
\put(667.16,647.67){\usebox{\plotpoint}}
\multiput(673,651)(17.270,11.513){0}{\usebox{\plotpoint}}
\multiput(679,655)(18.564,9.282){0}{\usebox{\plotpoint}}
\put(685.10,658.05){\usebox{\plotpoint}}
\multiput(691,661)(18.564,9.282){0}{\usebox{\plotpoint}}
\multiput(697,664)(18.564,9.282){0}{\usebox{\plotpoint}}
\put(703.69,667.29){\usebox{\plotpoint}}
\multiput(710,670)(19.690,6.563){0}{\usebox{\plotpoint}}
\multiput(716,672)(18.564,9.282){0}{\usebox{\plotpoint}}
\put(722.81,675.27){\usebox{\plotpoint}}
\multiput(728,677)(20.473,3.412){0}{\usebox{\plotpoint}}
\multiput(734,678)(19.690,6.563){0}{\usebox{\plotpoint}}
\put(742.85,680.41){\usebox{\plotpoint}}
\multiput(747,681)(20.473,3.412){0}{\usebox{\plotpoint}}
\multiput(753,682)(20.473,3.412){0}{\usebox{\plotpoint}}
\put(763.40,683.00){\usebox{\plotpoint}}
\multiput(765,683)(20.756,0.000){0}{\usebox{\plotpoint}}
\put(771,683){\usebox{\plotpoint}}
\put(220.00,135.00){\usebox{\plotpoint}}
\put(227.54,154.34){\usebox{\plotpoint}}
\multiput(229,158)(8.698,18.845){0}{\usebox{\plotpoint}}
\put(236.04,173.26){\usebox{\plotpoint}}
\put(245.50,191.71){\usebox{\plotpoint}}
\multiput(248,196)(9.939,18.221){0}{\usebox{\plotpoint}}
\put(255.56,209.87){\usebox{\plotpoint}}
\put(265.91,227.85){\usebox{\plotpoint}}
\multiput(266,228)(10.679,17.798){0}{\usebox{\plotpoint}}
\put(276.95,245.42){\usebox{\plotpoint}}
\multiput(278,247)(12.743,16.383){0}{\usebox{\plotpoint}}
\put(289.14,262.21){\usebox{\plotpoint}}
\multiput(291,265)(12.453,16.604){0}{\usebox{\plotpoint}}
\put(301.44,278.92){\usebox{\plotpoint}}
\multiput(303,281)(12.453,16.604){0}{\usebox{\plotpoint}}
\put(313.89,295.52){\usebox{\plotpoint}}
\multiput(315,297)(13.508,15.759){0}{\usebox{\plotpoint}}
\put(327.85,310.85){\usebox{\plotpoint}}
\multiput(328,311)(13.508,15.759){0}{\usebox{\plotpoint}}
\multiput(334,318)(13.508,15.759){0}{\usebox{\plotpoint}}
\put(341.37,326.60){\usebox{\plotpoint}}
\multiput(346,332)(13.508,15.759){0}{\usebox{\plotpoint}}
\put(355.13,342.13){\usebox{\plotpoint}}
\multiput(358,345)(14.676,14.676){0}{\usebox{\plotpoint}}
\put(369.80,356.80){\usebox{\plotpoint}}
\multiput(371,358)(14.676,14.676){0}{\usebox{\plotpoint}}
\multiput(377,364)(14.676,14.676){0}{\usebox{\plotpoint}}
\put(384.48,371.48){\usebox{\plotpoint}}
\multiput(389,376)(14.676,14.676){0}{\usebox{\plotpoint}}
\put(399.46,385.83){\usebox{\plotpoint}}
\multiput(402,388)(14.676,14.676){0}{\usebox{\plotpoint}}
\multiput(408,394)(14.676,14.676){0}{\usebox{\plotpoint}}
\put(414.31,400.31){\usebox{\plotpoint}}
\multiput(420,406)(14.676,14.676){0}{\usebox{\plotpoint}}
\put(429.25,414.71){\usebox{\plotpoint}}
\multiput(432,417)(15.759,13.508){0}{\usebox{\plotpoint}}
\multiput(439,423)(15.945,13.287){0}{\usebox{\plotpoint}}
\put(445.10,428.10){\usebox{\plotpoint}}
\multiput(451,434)(15.945,13.287){0}{\usebox{\plotpoint}}
\put(460.26,442.26){\usebox{\plotpoint}}
\multiput(463,445)(15.945,13.287){0}{\usebox{\plotpoint}}
\multiput(469,450)(15.945,13.287){0}{\usebox{\plotpoint}}
\put(476.02,455.73){\usebox{\plotpoint}}
\multiput(482,460)(14.676,14.676){0}{\usebox{\plotpoint}}
\put(491.78,469.15){\usebox{\plotpoint}}
\multiput(494,471)(15.945,13.287){0}{\usebox{\plotpoint}}
\multiput(500,476)(15.945,13.287){0}{\usebox{\plotpoint}}
\put(507.73,482.44){\usebox{\plotpoint}}
\multiput(512,486)(16.889,12.064){0}{\usebox{\plotpoint}}
\put(524.06,495.22){\usebox{\plotpoint}}
\multiput(525,496)(15.945,13.287){0}{\usebox{\plotpoint}}
\multiput(531,501)(15.945,13.287){0}{\usebox{\plotpoint}}
\put(540.01,508.51){\usebox{\plotpoint}}
\multiput(543,511)(15.945,13.287){0}{\usebox{\plotpoint}}
\multiput(549,516)(18.021,10.298){0}{\usebox{\plotpoint}}
\put(556.76,520.63){\usebox{\plotpoint}}
\multiput(562,525)(15.945,13.287){0}{\usebox{\plotpoint}}
\put(573.09,533.40){\usebox{\plotpoint}}
\multiput(574,534)(15.945,13.287){0}{\usebox{\plotpoint}}
\multiput(580,539)(17.270,11.513){0}{\usebox{\plotpoint}}
\put(589.78,545.70){\usebox{\plotpoint}}
\multiput(593,548)(17.270,11.513){0}{\usebox{\plotpoint}}
\multiput(599,552)(17.270,11.513){0}{\usebox{\plotpoint}}
\put(606.98,557.32){\usebox{\plotpoint}}
\multiput(611,560)(15.945,13.287){0}{\usebox{\plotpoint}}
\multiput(617,565)(17.270,11.513){0}{\usebox{\plotpoint}}
\put(623.80,569.40){\usebox{\plotpoint}}
\multiput(629,572)(18.021,10.298){0}{\usebox{\plotpoint}}
\put(641.73,579.82){\usebox{\plotpoint}}
\multiput(642,580)(17.270,11.513){0}{\usebox{\plotpoint}}
\multiput(648,584)(18.564,9.282){0}{\usebox{\plotpoint}}
\put(659.42,590.61){\usebox{\plotpoint}}
\multiput(660,591)(18.564,9.282){0}{\usebox{\plotpoint}}
\multiput(666,594)(19.077,8.176){0}{\usebox{\plotpoint}}
\put(678.12,599.56){\usebox{\plotpoint}}
\multiput(679,600)(18.564,9.282){0}{\usebox{\plotpoint}}
\multiput(685,603)(19.690,6.563){0}{\usebox{\plotpoint}}
\multiput(691,605)(18.564,9.282){0}{\usebox{\plotpoint}}
\put(697.03,608.01){\usebox{\plotpoint}}
\multiput(703,610)(19.957,5.702){0}{\usebox{\plotpoint}}
\multiput(710,612)(19.690,6.563){0}{\usebox{\plotpoint}}
\put(716.82,614.27){\usebox{\plotpoint}}
\multiput(722,616)(19.690,6.563){0}{\usebox{\plotpoint}}
\multiput(728,618)(20.473,3.412){0}{\usebox{\plotpoint}}
\put(736.85,619.47){\usebox{\plotpoint}}
\multiput(740,620)(20.547,2.935){0}{\usebox{\plotpoint}}
\multiput(747,621)(20.473,3.412){0}{\usebox{\plotpoint}}
\put(757.34,622.72){\usebox{\plotpoint}}
\multiput(759,623)(20.756,0.000){0}{\usebox{\plotpoint}}
\multiput(765,623)(20.756,0.000){0}{\usebox{\plotpoint}}
\put(771,623){\usebox{\plotpoint}}
\put(220.00,132.00){\usebox{\plotpoint}}
\put(227.88,151.20){\usebox{\plotpoint}}
\multiput(229,154)(8.698,18.845){0}{\usebox{\plotpoint}}
\put(236.43,170.11){\usebox{\plotpoint}}
\put(246.29,188.32){\usebox{\plotpoint}}
\multiput(248,191)(9.939,18.221){0}{\usebox{\plotpoint}}
\put(256.42,206.43){\usebox{\plotpoint}}
\multiput(260,213)(10.679,17.798){0}{\usebox{\plotpoint}}
\put(266.89,224.34){\usebox{\plotpoint}}
\multiput(272,232)(11.513,17.270){0}{\usebox{\plotpoint}}
\put(278.45,241.58){\usebox{\plotpoint}}
\multiput(285,250)(12.453,16.604){0}{\usebox{\plotpoint}}
\put(291.05,258.07){\usebox{\plotpoint}}
\multiput(297,266)(12.453,16.604){0}{\usebox{\plotpoint}}
\put(303.55,274.64){\usebox{\plotpoint}}
\multiput(309,281)(13.508,15.759){0}{\usebox{\plotpoint}}
\put(317.06,290.40){\usebox{\plotpoint}}
\multiput(321,295)(14.676,14.676){0}{\usebox{\plotpoint}}
\put(331.12,305.64){\usebox{\plotpoint}}
\multiput(334,309)(13.508,15.759){0}{\usebox{\plotpoint}}
\put(345.03,321.03){\usebox{\plotpoint}}
\multiput(346,322)(14.676,14.676){0}{\usebox{\plotpoint}}
\multiput(352,328)(13.508,15.759){0}{\usebox{\plotpoint}}
\put(359.27,336.09){\usebox{\plotpoint}}
\multiput(365,341)(14.676,14.676){0}{\usebox{\plotpoint}}
\put(374.34,350.34){\usebox{\plotpoint}}
\multiput(377,353)(15.945,13.287){0}{\usebox{\plotpoint}}
\multiput(383,358)(14.676,14.676){0}{\usebox{\plotpoint}}
\put(389.50,364.50){\usebox{\plotpoint}}
\multiput(395,370)(16.889,12.064){0}{\usebox{\plotpoint}}
\put(405.09,378.09){\usebox{\plotpoint}}
\multiput(408,381)(15.945,13.287){0}{\usebox{\plotpoint}}
\multiput(414,386)(14.676,14.676){0}{\usebox{\plotpoint}}
\put(420.26,392.22){\usebox{\plotpoint}}
\multiput(426,397)(14.676,14.676){0}{\usebox{\plotpoint}}
\put(435.91,405.79){\usebox{\plotpoint}}
\multiput(439,408)(15.945,13.287){0}{\usebox{\plotpoint}}
\multiput(445,413)(15.945,13.287){0}{\usebox{\plotpoint}}
\put(452.03,418.86){\usebox{\plotpoint}}
\multiput(457,423)(15.945,13.287){0}{\usebox{\plotpoint}}
\put(467.97,432.14){\usebox{\plotpoint}}
\multiput(469,433)(15.945,13.287){0}{\usebox{\plotpoint}}
\multiput(475,438)(16.889,12.064){0}{\usebox{\plotpoint}}
\put(484.31,444.92){\usebox{\plotpoint}}
\multiput(488,448)(15.945,13.287){0}{\usebox{\plotpoint}}
\multiput(494,453)(15.945,13.287){0}{\usebox{\plotpoint}}
\put(500.27,458.18){\usebox{\plotpoint}}
\multiput(506,462)(15.945,13.287){0}{\usebox{\plotpoint}}
\put(516.93,470.52){\usebox{\plotpoint}}
\multiput(519,472)(17.270,11.513){0}{\usebox{\plotpoint}}
\multiput(525,476)(15.945,13.287){0}{\usebox{\plotpoint}}
\put(533.66,482.77){\usebox{\plotpoint}}
\multiput(537,485)(15.945,13.287){0}{\usebox{\plotpoint}}
\multiput(543,490)(17.270,11.513){0}{\usebox{\plotpoint}}
\put(550.40,495.00){\usebox{\plotpoint}}
\multiput(556,499)(17.270,11.513){0}{\usebox{\plotpoint}}
\put(567.54,506.69){\usebox{\plotpoint}}
\multiput(568,507)(17.270,11.513){0}{\usebox{\plotpoint}}
\multiput(574,511)(17.270,11.513){0}{\usebox{\plotpoint}}
\put(584.44,518.70){\usebox{\plotpoint}}
\multiput(586,520)(18.021,10.298){0}{\usebox{\plotpoint}}
\multiput(593,524)(18.564,9.282){0}{\usebox{\plotpoint}}
\put(602.29,529.20){\usebox{\plotpoint}}
\multiput(605,531)(17.270,11.513){0}{\usebox{\plotpoint}}
\multiput(611,535)(17.270,11.513){0}{\usebox{\plotpoint}}
\put(619.75,540.38){\usebox{\plotpoint}}
\multiput(623,542)(17.270,11.513){0}{\usebox{\plotpoint}}
\multiput(629,546)(19.077,8.176){0}{\usebox{\plotpoint}}
\put(637.91,550.28){\usebox{\plotpoint}}
\multiput(642,553)(18.564,9.282){0}{\usebox{\plotpoint}}
\multiput(648,556)(18.564,9.282){0}{\usebox{\plotpoint}}
\put(656.17,560.09){\usebox{\plotpoint}}
\multiput(660,562)(18.564,9.282){0}{\usebox{\plotpoint}}
\multiput(666,565)(19.077,8.176){0}{\usebox{\plotpoint}}
\put(674.92,568.96){\usebox{\plotpoint}}
\multiput(679,571)(19.690,6.563){0}{\usebox{\plotpoint}}
\multiput(685,573)(19.690,6.563){0}{\usebox{\plotpoint}}
\put(694.18,576.59){\usebox{\plotpoint}}
\multiput(697,578)(19.690,6.563){0}{\usebox{\plotpoint}}
\multiput(703,580)(19.957,5.702){0}{\usebox{\plotpoint}}
\put(713.94,582.66){\usebox{\plotpoint}}
\multiput(716,583)(19.690,6.563){0}{\usebox{\plotpoint}}
\multiput(722,585)(20.473,3.412){0}{\usebox{\plotpoint}}
\put(733.94,587.98){\usebox{\plotpoint}}
\multiput(734,588)(20.473,3.412){0}{\usebox{\plotpoint}}
\multiput(740,589)(20.756,0.000){0}{\usebox{\plotpoint}}
\multiput(747,589)(20.473,3.412){0}{\usebox{\plotpoint}}
\put(754.50,590.25){\usebox{\plotpoint}}
\multiput(759,591)(20.756,0.000){0}{\usebox{\plotpoint}}
\multiput(765,591)(20.756,0.000){0}{\usebox{\plotpoint}}
\put(771,591){\usebox{\plotpoint}}
\end{picture}

%% file: phasefig5.tex
\setlength{\unitlength}{0.240900pt}
\ifx\plotpoint\undefined\newsavebox{\plotpoint}\fi
\sbox{\plotpoint}{\rule[-0.200pt]{0.400pt}{0.400pt}}%
\begin{picture}(1500,900)(0,0)
\font\gnuplot=cmr10 at 10pt
\gnuplot
\sbox{\plotpoint}{\rule[-0.200pt]{0.400pt}{0.400pt}}%
\put(220.0,113.0){\rule[-0.200pt]{4.818pt}{0.400pt}}
\put(198,113){\makebox(0,0)[r]{0.1}}
\put(1416.0,113.0){\rule[-0.200pt]{4.818pt}{0.400pt}}
\put(220.0,343.0){\rule[-0.200pt]{2.409pt}{0.400pt}}
\put(1426.0,343.0){\rule[-0.200pt]{2.409pt}{0.400pt}}
\put(220.0,478.0){\rule[-0.200pt]{2.409pt}{0.400pt}}
\put(1426.0,478.0){\rule[-0.200pt]{2.409pt}{0.400pt}}
\put(220.0,573.0){\rule[-0.200pt]{2.409pt}{0.400pt}}
\put(1426.0,573.0){\rule[-0.200pt]{2.409pt}{0.400pt}}
\put(220.0,647.0){\rule[-0.200pt]{2.409pt}{0.400pt}}
\put(1426.0,647.0){\rule[-0.200pt]{2.409pt}{0.400pt}}
\put(220.0,708.0){\rule[-0.200pt]{2.409pt}{0.400pt}}
\put(1426.0,708.0){\rule[-0.200pt]{2.409pt}{0.400pt}}
\put(220.0,759.0){\rule[-0.200pt]{2.409pt}{0.400pt}}
\put(1426.0,759.0){\rule[-0.200pt]{2.409pt}{0.400pt}}
\put(220.0,803.0){\rule[-0.200pt]{2.409pt}{0.400pt}}
\put(1426.0,803.0){\rule[-0.200pt]{2.409pt}{0.400pt}}
\put(220.0,842.0){\rule[-0.200pt]{2.409pt}{0.400pt}}
\put(1426.0,842.0){\rule[-0.200pt]{2.409pt}{0.400pt}}
\put(220.0,877.0){\rule[-0.200pt]{4.818pt}{0.400pt}}
\put(198,877){\makebox(0,0)[r]{1}}
\put(1416.0,877.0){\rule[-0.200pt]{4.818pt}{0.400pt}}
\put(220.0,113.0){\rule[-0.200pt]{0.400pt}{4.818pt}}
\put(220,68){\makebox(0,0){1}}
\put(220.0,857.0){\rule[-0.200pt]{0.400pt}{4.818pt}}
\put(463.0,113.0){\rule[-0.200pt]{0.400pt}{4.818pt}}
\put(463,68){\makebox(0,0){1.02}}
\put(463.0,857.0){\rule[-0.200pt]{0.400pt}{4.818pt}}
\put(706.0,113.0){\rule[-0.200pt]{0.400pt}{4.818pt}}
\put(706,68){\makebox(0,0){1.04}}
\put(706.0,857.0){\rule[-0.200pt]{0.400pt}{4.818pt}}
\put(950.0,113.0){\rule[-0.200pt]{0.400pt}{4.818pt}}
\put(950,68){\makebox(0,0){1.06}}
\put(950.0,857.0){\rule[-0.200pt]{0.400pt}{4.818pt}}
\put(1193.0,113.0){\rule[-0.200pt]{0.400pt}{4.818pt}}
\put(1193,68){\makebox(0,0){1.08}}
\put(1193.0,857.0){\rule[-0.200pt]{0.400pt}{4.818pt}}
\put(1436.0,113.0){\rule[-0.200pt]{0.400pt}{4.818pt}}
\put(1436,68){\makebox(0,0){1.1}}
\put(1436.0,857.0){\rule[-0.200pt]{0.400pt}{4.818pt}}
\put(220.0,113.0){\rule[-0.200pt]{292.934pt}{0.400pt}}
\put(1436.0,113.0){\rule[-0.200pt]{0.400pt}{184.048pt}}
\put(220.0,877.0){\rule[-0.200pt]{292.934pt}{0.400pt}}
\put(45,495){\makebox(0,0){$\sigma$}}
\put(828,23){\makebox(0,0){$\beta$}}
\put(220.0,113.0){\rule[-0.200pt]{0.400pt}{184.048pt}}
\put(1314,803){\makebox(0,0)[r]{fluctuating lattice}}
\put(1358,803){\raisebox{-.8pt}{\makebox(0,0){$\Box$}}}
\put(731,547){\raisebox{-.8pt}{\makebox(0,0){$\Box$}}}
\put(950,499){\raisebox{-.8pt}{\makebox(0,0){$\Box$}}}
\put(1132,417){\raisebox{-.8pt}{\makebox(0,0){$\Box$}}}
\put(1193,375){\raisebox{-.8pt}{\makebox(0,0){$\Box$}}}
\put(731.0,519.0){\rule[-0.200pt]{0.400pt}{13.009pt}}
\put(721.0,519.0){\rule[-0.200pt]{4.818pt}{0.400pt}}
\put(721.0,573.0){\rule[-0.200pt]{4.818pt}{0.400pt}}
\put(950.0,478.0){\rule[-0.200pt]{0.400pt}{9.877pt}}
\put(940.0,478.0){\rule[-0.200pt]{4.818pt}{0.400pt}}
\put(940.0,519.0){\rule[-0.200pt]{4.818pt}{0.400pt}}
\put(1132.0,375.0){\rule[-0.200pt]{0.400pt}{19.272pt}}
\put(1122.0,375.0){\rule[-0.200pt]{4.818pt}{0.400pt}}
\put(1122.0,455.0){\rule[-0.200pt]{4.818pt}{0.400pt}}
\put(1193.0,343.0){\rule[-0.200pt]{0.400pt}{14.454pt}}
\put(1183.0,343.0){\rule[-0.200pt]{4.818pt}{0.400pt}}
\put(1183.0,403.0){\rule[-0.200pt]{4.818pt}{0.400pt}}
\sbox{\plotpoint}{\rule[-0.500pt]{1.000pt}{1.000pt}}%
\put(220,704){\usebox{\plotpoint}}
\multiput(220,704)(19.848,-6.072){31}{\usebox{\plotpoint}}
\multiput(828,518)(19.848,-6.072){31}{\usebox{\plotpoint}}
\put(1436,332){\usebox{\plotpoint}}
\sbox{\plotpoint}{\rule[-0.200pt]{0.400pt}{0.400pt}}%
\put(1314,758){\makebox(0,0)[r]{pure gauge}}
\put(1358,758){\circle{24}}
\put(463,565){\circle{24}}
\put(524,565){\circle{24}}
\put(585,499){\circle{24}}
\put(646,455){\circle{24}}
\put(706,430){\circle{24}}
\put(463.0,556.0){\rule[-0.200pt]{0.400pt}{4.095pt}}
\put(453.0,556.0){\rule[-0.200pt]{4.818pt}{0.400pt}}
\put(453.0,573.0){\rule[-0.200pt]{4.818pt}{0.400pt}}
\put(524.0,538.0){\rule[-0.200pt]{0.400pt}{12.286pt}}
\put(514.0,538.0){\rule[-0.200pt]{4.818pt}{0.400pt}}
\put(514.0,589.0){\rule[-0.200pt]{4.818pt}{0.400pt}}
\put(585.0,478.0){\rule[-0.200pt]{0.400pt}{9.877pt}}
\put(575.0,478.0){\rule[-0.200pt]{4.818pt}{0.400pt}}
\put(575.0,519.0){\rule[-0.200pt]{4.818pt}{0.400pt}}
\put(646.0,430.0){\rule[-0.200pt]{0.400pt}{11.563pt}}
\put(636.0,430.0){\rule[-0.200pt]{4.818pt}{0.400pt}}
\put(636.0,478.0){\rule[-0.200pt]{4.818pt}{0.400pt}}
\put(706.0,417.0){\rule[-0.200pt]{0.400pt}{6.263pt}}
\put(696.0,417.0){\rule[-0.200pt]{4.818pt}{0.400pt}}
\put(696.0,443.0){\rule[-0.200pt]{4.818pt}{0.400pt}}
\sbox{\plotpoint}{\rule[-0.500pt]{1.000pt}{1.000pt}}%
\put(220,619){\usebox{\plotpoint}}
\multiput(220,619)(19.848,-6.072){31}{\usebox{\plotpoint}}
\multiput(828,433)(19.848,-6.072){31}{\usebox{\plotpoint}}
\put(1436,247){\usebox{\plotpoint}}
\end{picture}